\documentclass[aps, amsmath, amssymb, floatfix, pre, showpacs, twocolumn,
letterpaper]{revtex4}  \newcommand{\Figurescale}{1.0}%
\usepackage[american]{babel}%
\usepackage{graphicx}%
\usepackage{prettyref}%
\newrefformat{apx}{Appendix~\ref{#1}}%
\newrefformat{fig}{Fig.~\ref{#1}}%
\newrefformat{sec}{Sec.~\ref{#1}}%
\newrefformat{subsec}{Sec.~\ref{#1}}%
\newrefformat{eqn}{Eq.~(\ref{#1})}%
\newcommand{\Lang}{\left \langle} \newcommand{\Rang}{\right \rangle}%
\newcommand{\Lbar}{\left |{}} \newcommand{\Rbar}{\right|}%
\newcommand{\Lbkt}{\left[{}} \newcommand{\Rbkt}{\right]}%
\newcommand{\Lpar}{\left({}} \newcommand{\Rpar}{\right)}%
\newcommand{\Rf}{\mathbf{r}^\mathrm{f}}%
\newcommand{\Rt}{\mathbf{r}^\mathrm{t}}%
\newcommand{\Phif}{\phi_\mathrm{f}}%
\newcommand{\Phim}{\phi_\mathrm{m}}%
\newcommand{\Fc}{F_\mathrm{c}}%
\newcommand{\Fs}{F_\mathrm{s}}%
\newcommand{\Gs}{G_\mathrm{s}}%
\newcommand{\Nf}{N_\mathrm{f}}%
\newcommand{\Nm}{N_\mathrm{m}}%
\newcommand{\Sb}{S_\mathrm{b}}%
\newcommand{\Sc}{S_\mathrm{c}}%
\newcommand{\Tc}{t_\mathrm{c}}%
\newcommand{\Td}{t_\mathrm{d}}%
\newcommand{\Te}{t_\mathrm{e}}%
\newcommand{\Ts}{t_\mathrm{s}}%
\newcommand{\Tt}{t_\mathrm{t}}%
\newcommand{\Tts}{\tilde t_\mathrm{s}}%
\newcommand{\Fts}{\tilde F_\mathrm{s}}%
\newcommand{\Drs}{\delta r^2}%
\newcommand{\Drmk}{\delta\rho_\mathbf{-k}}%
\newcommand{\Drpk}{\delta\rho_\mathbf{k}}%
\newcommand{\Rmk}{\rho_\mathbf{-k}}%
\newcommand{\Rpk}{\rho_\mathbf{k}}%
\newcommand{\Rtmk}{\rho_\mathbf{-k}^\mathrm{t}}%
\newcommand{\Rtpk}{\rho_\mathbf{k}^\mathrm{t}}%
\newcommand{\Cf}{cf.}%
\newcommand{\Ie}{i.e.}%
\usepackage{color}%
\definecolor{changed}{rgb}{0., 0., 1.}%
\begin{document}%
\newlength{\Figurewidth}%
\setlength{\Figurewidth}{\Figurescale\columnwidth}%
\title{Impact of random obstacles on the dynamics
of a dense colloidal fluid}%

\author{Jan Kurzidim}%
\author{Daniele Coslovich}%
\author{Gerhard Kahl}%

\affiliation{Institut f\"u{}r Theoretische Physik and CMS, Technische
Universit\"a{}t Wien, Wiedner Hauptstra\ss{}e 8-10, 1040 Wien,
Austria}%

\date{\today}%

\begin{abstract}%

Using molecular dynamics simulations we study the slow dynamics of a
colloidal fluid annealed within a matrix of obstacles quenched from
an equilibrated colloidal fluid. We choose all particles to be of
the same size and to interact as hard spheres, thus retaining all
features of the porous confinement while limiting the control
parameters to the packing fraction of the matrix,~$\Phim$, and that
of the fluid,~$\Phif$. We conduct detailed investigations on several
dynamic properties, including the tagged-particle and collective
intermediate scattering functions, the mean-squared displacement,
and the van Hove function. We show the confining obstacles to
profoundly impact the relaxation pattern of various quantifiers
pertinent to the fluid. Varying the type of quantifier
(tagged-particle or collective) as well as $\Phim$ and $\Phif$, we
unveil both discontinuous and continuous arrest scenarios.
Furthermore, we discover subdiffusive behavior and demonstrate its
close connection to the matrix structure. Our findings partly
confirm the various predictions of a recent extension of
mode-coupling theory to the quenched-annealed protocol.

\end{abstract}%


\maketitle
\section{Introduction}%
\label{sec:Introduction}%
Fluids that are brought into contact with
a disordered porous matrix
drastically change their physical properties. Over the past two or
three decades numerous experimental and theoretical investigations
have been dedicated to studying this phenomenon~\cite{gelb1999,
rosinberg1999, mckenna2000, mckenna2003, alcoutlabi2005, mckenna2007}. The
remarkable efforts that have been undertaken to obtain a deeper
understanding of the underlying mechanisms are not only motivated by
academic interest. Fluids confined in disordered porous environments
play a key role in a wide spectrum of applied problems ranging from
technological applications over chemical engineering to biophysical
systems. Thus, a deeper understanding of \emph{why} fluids change
their properties under such external conditions is of great practical
relevance.

From a theoretical viewpoint, a quantitative description of such
systems is a formidable challenge for several reasons. For one, it is
difficult to reliably and realistically represent the matrices found
in experiments or technological applications (such as aerogels or
vycor). Having chosen some model for the confinement, an even more
demanding task is to formulate a theoretical framework that is able to
appropriately describe the interplay of connectivity and confinement
of the pores pertinent to the matrix. Even after reducing such system
to the so-called quenched-annealed (QA) model, where for simplicity
the matrix is assumed to be an instantaneously-frozen configuration of
an equilibrated fluid, an appropriate theory is highly intricate.
Madden and Glandt~\cite{madden1988, madden1992}, and later Given
and Stell~\cite{given1992a, given1992b, given1994}, derived a
formalism based on statistical mechanics that considers the system to
be a peculiar mixture with the matrix being one of the components.
Within their framework, Given and Stell derived Ornstein-Zernike-type
integral equations (the so-called replica Ornstein-Zernike equations,
ROZ) that---in combination with a suitable closure relation---allow to
determine the static correlators and the thermodynamic properties of a
QA system~\cite{rosinberg1994, kierlik1995, kierlik1997,
paschinger2000, paschinger2001}. Their formalism is based on a
twofold thermodynamic averaging procedure: one over all possible fluid
configurations for a fixed matrix realization, the other one over all
possible matrix configurations. This double-averaging procedure is
adapted to the letter in computer simulation studies of fluids
confined in disordered porous materials; in terms of computing time
this is a demanding enterprise.

Until few years ago most theoretical and simulation studies focused on
the \emph{static} properties of QA systems such as structure and
phase behavior. In contrast, comparatively little effort had been
devoted to the \emph{dynamic} properties. The main obstacles
inhibiting such investigations were, on one side, the lack of a
suitable theoretical framework that would have allowed to reliably
evaluate the dynamic correlation functions of the fluid particles,
and, on the other hand, limitations in computational power for
simulations when explicitly performing the above-mentioned
double-averaging procedure. To the best of our knowledge, the first
simulation studies dedicated to the dynamic properties of fluids in
disordered porous confinement were performed by Gallo and Rovere~\cite
{gallo2002, gallo2003a, gallo2003b}, followed shortly after by
Kim~\cite{kim2003}. In this context it is worth mentioning similar
investigations on the Lorentz gas model~\cite{hofling2006} and on
binary mixtures in which particles are characterized by a large
disparity in size~\cite{voigtmann2009, moreno2006} or mass~\cite{fenz2009},
\Ie, systems that can be viewed to be closely related to
QA systems.

Over the past years, major breakthroughs have been achieved in
describing theoretically the dynamic properties of fluids in confined
in porous media. In particular, use was made of mode-coupling theory
(MCT)~\cite{krakoviack2005, krakoviack2007, krakoviack2009} and of
the self-consistent generalized Langevin equation (SCGLE) theory~\cite
{chavez2008}, both of which rely on inferring the dynamics of a
system from its static structure. MCT has played over the last decades
a central role in describing the dynamic properties of fluids~\cite{gotze1999},
including in particular the glass transition. Krakoviack
succeeded in deriving MCT-type equations for the dynamic correlation
functions, which require as an input the static correlation functions
of the annealed fluid, which in turn are obtained from the ROZ
equations. Subsequently, Krakoviack solved these equations for the
specific case of a hard-sphere fluid confined in a disordered matrix
of hard-sphere particles of equal size and evaluated single-particle
and collective correlation functions of the system. Based on this
knowledge, a kinetic diagram was traced out in the parameter space
spanned by the packing fraction of the fluid and that of the matrix,
indicating the regions in which the system is in an arrested or in a
nonarrested state. The resulting kinetic diagram is very rich and
quite surprising: the regions of arrested and nonarrested states are
separated by a line along which at small matrix packing fractions
type-B transitions and at intermediate matrix packing fractions type-A
transitions occur. As a peculiar feature of this line, a re-entrant
glass transition is predicted for intermediate matrix and small fluid
packing fractions. Additionally, in the region of nonarrested states a
continuous diffusion-localization transition is observed. However, we
emphasize that the combined ROZ+MCT framework predicts \emph{ideal}
transitions, whereas in experiment or simulation transitions will
always be smeared out.

Throughout this contribution we employ the type-A/type-B terminology
that is conventionally used to characterize the behavior of a dynamic
correlator as an ideal transition occurs. This terminology refers to
the decay pattern of the correlator as well as to the behavior of its
long-time limit upon crossing a dynamic transition. In a ``type-A''
transition a correlator relaxes in a single step, and its long-time
limit may assume arbitrarily small nonzero values. In a transition of
``type-B,'' on the other hand, a correlator decays in a distinct
two-step pattern; upon crossing the transition the second decay step
is delayed to infinity and the long-time limit of the correlator jumps
from zero to a nonzero value. The type-B behavior is well-known from
bulk glass formers~\cite{gotze1992} and is usually attributed to a
``cage effect'' imposed upon fluid particles by their neighbors.

This remarkably rich phase behavior has motivated simulation studies
which---as a consequence of the considerable increase in available
computation power---have meanwhile come within reach. Short accounts
have been given in~\cite{kurzidim2009} and~\cite{kim2009}. In
this contribution we elaborate on our investigations on the dynamic
behavior of a fluid confined in a disordered porous matrix within the
QA picture. For simplicity, and for congruence with the theoretical
studies \cite{krakoviack2005, krakoviack2007, krakoviack2009}, we
consider a fluid of hard spheres confined in a disordered matrix
formed by hard spheres of the same size. Investigations are based on
event-driven molecular dynamics simulations. Within the
two-dimensional parameter space we focus mainly on three
representative pathways to put the theoretical predictions to a
thorough test: (i) a path at constant low matrix packing fraction,
probing the existence of the continuous type-B transition, (ii) a path
at constant intermediate matrix packing fraction, probing the
discontinuous type-A transition and the re-entrant scenario, and (iii)
a path at intermediate, constant fluid packing fraction, probing the
diffusion-localization transition. Results are summarized in kinetic
diagrams, where we consider---complementing our previous
communication---different criteria to define dynamical arrest. We
compare our simulation results with the predictions of MCT, which,
however, is not straightforward due to the ideal nature of the latter.

The paper is organized as follows: in \prettyref{sec:Model.and.Methods} we
present the model and our simulation
method, transferring the rather lengthy description of the system
setup to \prettyref{apx:System.setup.details}. Results are compiled
in \prettyref{sec:Results}: we start with the static structure of
the system and then discuss various methods of how to define slowing
down and the ensuing kinetic diagrams. In the subsequent three
subsections we discuss the intermediate scattering functions and the
mean-squared displacement along the three aforementioned paths through
parameter space. The final subsection of \prettyref{sec:Results}
shall deal with the van Hove correlation function in order to discuss
trapping. In \prettyref{sec:Discussion} we extend the analysis of
the intermediate scattering functions to discuss the arrest scenarios
in more detail. The paper closes with concluding remarks alongside
with an outlook on future work required to settle questions yet open.
The aforementioned \prettyref{apx:System.setup.details} is
succeeded by \prettyref{apx:Finite-size.effects} in which we
elaborate on the issue of the effects of system size on the dynamical
correlation functions.
\section{Model and Methods}%
\label{sec:Model.and.Methods}%
\begin{figure}%
\includegraphics* [width = \Figurewidth]{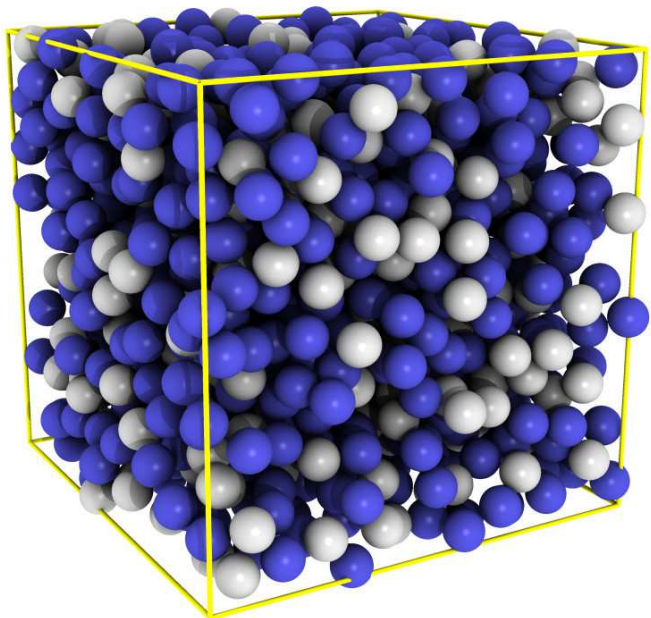}%
\caption{\label{fig:snapshot}%
(Color online) Snapshot of a quenched-annealed mixture of hard
spheres at $\Phim = 0.25$ and $\Phif = 0.10$. Dark (blue): matrix
particles, light (gray): fluid particles.}%
\end{figure}%

For the purpose of this work we opted for the ``quenched-annealed''
(QA) protocol since it is well-defined, well-tractable by theory, and
involves only a small number of parameters. In this protocol both the
porous confinement and the confined fluid are represented by a set of
particles, conventionally called ``matrix'' and ``fluid'' particles
or---borrowing from the typology of metallurgy---as ``quenched'' and
``annealed'' component. In this, QA systems bear resemblance to
mixtures; however, the matrix particles are pinned at particular
positions and act as obstacles for the fluid particles. In this work,
we exclusively consider systems containing one matrix and one fluid
species; quantities pertaining to those species will be designated
with the subscripts ``m'' and ``f,'' respectively.

\newcommand{\FootMatrixSetup}{\footnote{More detailed, the procedure to set
up a QA matrix reads as follows:
(1) define an interaction between the matrix particles, (2) find
some allowed configuration for a system containing just the matrix
particles, (3) equilibrate that system, (4) take a snapshot at some
instant of time, (5) use the particle positions therein as obstacle
positions.}}%

Adopting the concept of immobile obstacle particles, there are still
numerous schemes for their arrangement~\cite{zhang2000, zhao2006,
kim2009, fenz2009, gallo2009}. The QA protocol specifies the
positions of the matrix particles to be frozen from an equilibrium
one-component fluid; the fluid particles merely move within this
matrix and do not exert influence upon it~\FootMatrixSetup. Notably,
the fluid particles may also dwell in ``traps''---finite spatial
domains entirely bounded by infinite potential walls formed by the
confining matrix.

Since for this work we are interested in a minimal approach to the
dynamics of a fluid moving in disordered confinement, we decided to
exclusively employ hard-sphere (HS) interactions among the particles.
This deprives the system of any inherent energy scale, rendering the
only relevant system attributes to be of geometrical nature. We
further reduced the parameter space by requiring all particles
involved to be of the same (monodisperse) diameter~$\sigma$. This
introduces $\sigma$ as the unit of length used throughout this study,
and leaves the system to be governed by mere packing fractions, namely
the matrix packing fraction $\Phim$ and the fluid packing fraction
$\Phif$. As mentioned before, this particular choice of system was
examined in~\cite{krakoviack2005, krakoviack2007, krakoviack2009}.
A snapshot of a typical HS-QA system at large matrix density is shown
in \prettyref{fig:snapshot}.

In this work we study QA systems utilizing molecular simulations.
Since we are primarily interested in dynamical features, we applied
molecular dynamics (MD) simulations to track the physical time
evolution of a system. All MD computations were performed using the
event-driven algorithm described in \cite{alder1959}; modifications
to that algorithm were essentially limited to fixing particles in
space in order to adapt to the QA protocol. (Notably, this adjustment
invalidates the conservation of momentum and angular momentum.) As is
conventional, we employed periodic (toroidal) boundary conditions and
the minimum-image convention in order to mimic an infinite system.

In addition to the usual thermodynamic averaging, $\langle \mathstrut
\cdots \rangle$\,, the QA protocol imposes another averaging procedure
over \emph{disorder}, $\overline{\mathstrut \cdots}$\,, so as to
restore homogeneity and isotropy. Depending on the state point $(\Phim, \Phif)$
and the number of particles in the system, we averaged
our data over at least ten matrix configurations; for selected state
points, we increased that number to as much as $40$.

A nontrivial task is to find an initial system configuration given
some state point. In this work, we are interested in systems in which
the $\Nf$ fluid particles comprise most of the space that is left
accessible to them by the $\Nm$ immobile matrix particles---\Ie, \emph
{high-density} systems. Generating instances of the porous
confinement is effortless, since for all state points of interest,
$\Phim$ is situated well within the fluid regime when used for a
one-component system of mobile particles. The problem rests in
determining permissible initial positions for the $\Nf$ fluid
particles. Creating an amorphous high-density configuration of bulk
monodisperse hard-sphere particles already represents an interesting
task, with various elaborate algorithms having been developed to
accomplish this task~\cite{bennett1972, jodrey1985}. However, none
of these algorithms are suited for adaption to the setup of HS-QA
systems. This unfortunate finding prompted us to devise a custom
routine which is described in detail in \prettyref{apx:System.setup.details}.

After successful setup of a system instance, the simulation time was
set to $t = 0$ and an attempt to equilibrate the system using the MD
routine was made. The decision whether or not the particular system
realization was considered equilibrated was based on the mean-squared
displacement

\begin{equation} \label{eqn:def_Drst}%
\Drs(t) = \overline{\Lang \Lbar \Rt(t)-\Rt(0) \Rbar^2 \Rang}%
\end{equation}%

\noindent where $\Rt(t)$ is the location of a tagged (superscript
``$\mathrm{t}$'') fluid particle at time $t$. Note that the above
definition is the one used in \prettyref{sec:Results} and hence
involves an average over matrix configurations; of course this average
has to be omitted when using $\Drs(t)$ to characterize a single system
realization. Also, note that the notion of a ``tagged particle'' does
\emph{not} imply the presence of an additional tracer particle.

We defined a system realization to be in equilibrium if $\Drs(t_*) >%
\Drs_* = 100$ was fulfilled. We allowed at least $t_* = 30\,000$ time
units for the system to equilibrate; for certain state points we
extended $t_*$ to as much as the tenfold value. The conventional unit
of time $\tau = \sqrt{m \sigma^2 / k_\mathrm{B}T}$ used throughout
this work is based upon the mass $m$ of the fluid particles, their
diameter $\sigma$ (see above), and their temperature $T$~\cite{allen1987}.

\newcommand{\FootParticleNumber}{\footnote{Increases or decreases of $\Nm$
and/or $\Nf$ to accurately
approximate the desired fraction $\Nm/\Nf$ resulted in slight
deviations from the standard $N = 1000$.}}%

Data were only compiled after equilibration was attempted, using the
final configuration of the equilibration run as initial configuration
of the data run. Generally, the final time for data collection, $\Td$,
was chosen to be equal to the final time of the equilibration attempt,
$\Te$; in particular in some nonequilibrated systems we also used $\Td
> \Te$ to investigate particular features of the system. For most
state points, we chose $N = \Nm{+}\Nf \simeq 1000$ to be the total
particle number~\FootParticleNumber. In order to retain reasonable
statistics even close to $\Phif = 0$, we increased $N$ to as much as
$12\,000$ for elevated $\Phim$, such that $\Nf \geq 50$ for all state
points.
\section{Results}%
\label{sec:Results}%
\subsection{Static structure}%
\label{subsec:Static.structure}%
\begin{figure}%
\includegraphics* [width = \Figurewidth]{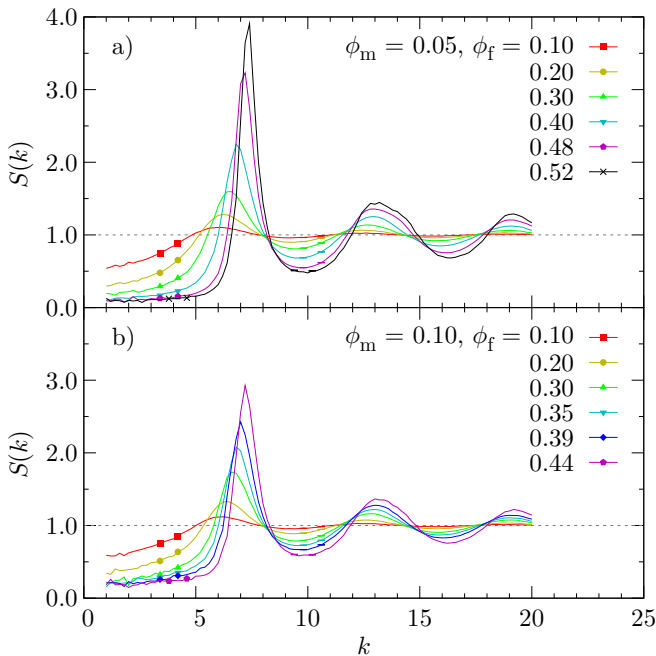}%
\caption{\label{fig:static/sk}%
(Color online) Fluid-fluid static structure factor $S(k)$ for a
series of $\Phif$ values at constant $\Phim$. (a) $\Phim = 0.05$,
(b) $\Phim = 0.10$. Error bars represent one standard deviation of
the mean for different system realizations.}%
\end{figure}%
\begin{figure}%
\includegraphics* [width = \Figurewidth]{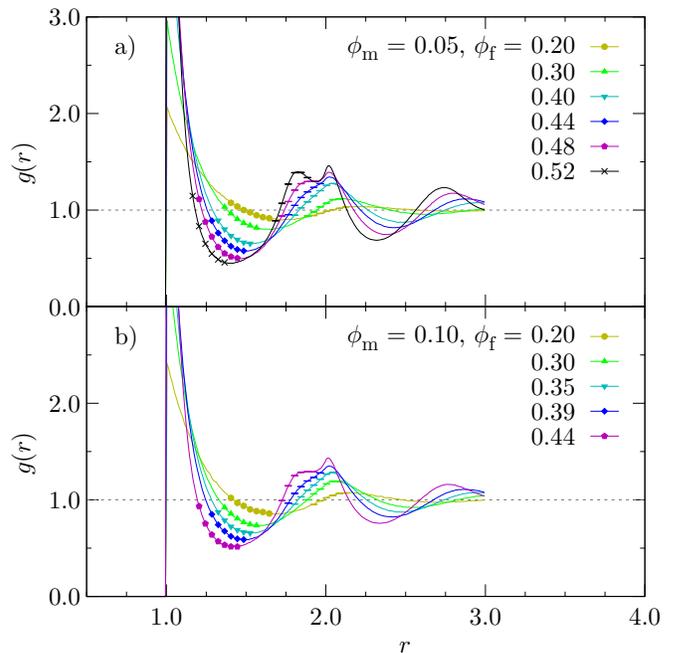}%
\caption{\label{fig:static/gr}%
(Color online) Fluid-fluid radial distribution function $g(r)$ for
a series of $\Phif$ values at constant $\Phim$. (a) $\Phim =
0.05$, (b) $\Phim = 0.10$. Error bars: see \prettyref{fig:static/sk}}%
\end{figure}%

The most obvious properties to inspect in our systems are simple
static structural properties; this investigation was already carried
out in great detail in~\cite{lomba1993}. In the present context
static properties are of use for a different reason: Since the
particles were chosen to be of monodisperse diameter~$\sigma$,
crystallization might take place in the case of dilute matrices.
Inspection of the radial distribution function $g(r)$ and the static
structure factor $S(k)$ provides a means to determine whether or not
this is the case. In fact, it is sufficient to investigate fluid-fluid
correlations, $g_\mathrm{ff}(r)$ and $S_\mathrm{ff}(k)$, since only
for $\Phif \gg \Phim$ systems are prone to crystallization. Therefore,
the index ``$\mathrm{ff}$'' will henceforth be dropped.

We first consider the static structure factor

\begin{equation} \label{eqn:def_Sk}
S(k) = \frac{4\pi}{\Nf}~
\overline{\Lang
\sum_i^{\Nf} \sum_j^{\Nf}%
\frac{r_{ij}}{k} \sin \Lpar k\,r_{ij} \Rpar
\Rang}%
\end{equation}%

\noindent where $r_{ij} = |\Rf_i-\Rf_j|$, and $\Rf_i$ denotes the
location of fluid particle $i$. \prettyref{fig:static/sk} displays
this function for moderate ($\Phim = 0.1$) and for low ($\Phim =
0.05$) matrix packing fractions, \Ie, for systems particularly prone
to crystallization. Clearly, all features of $S(k)$ evolve smoothly
upon increasing $\Phif$, suggesting that no phase transition takes
place. Moreover, the absence of sharp peaks for all state points
indicates that no simple crystalline long-range order is present.

Related arguments hold for the radial distribution function

\begin{equation} \label{eqn:def_gr}
g(r) = \frac{V}{\Nf^2}~
\overline{\Lang
\mathop{\sum\sum}_{i \ne j}^{\Nf} \delta \Lpar r - r_{ij} \Rpar
\Rang}%
\end{equation}%

\noindent where $V$ is the system volume. \prettyref{fig:static/gr}
shows this function for the same matrix packing fractions as above
($\Phim \in \{0.1, 0.05\}$). As for $S(k)$, the features of $g(r)$
develop continuously, which again supports the notion that a phase
transition is absent.

Additional peaks in $g(r)$ at specific positions would be a hallmark
of crystallization. For fcc- or bcc-like short-range order, the most
pronounced additional peaks would be located at $r_\mathrm{fcc} = \Phi
\sqrt{2}$ or $r_\mathrm{bcc} = \Phi \sqrt{3}$, where $\Phi = \{%
(\Phim+\Phif)/\phi_\mathrm{max} \}^{1/3}$ and $\phi_\mathrm{max} = \pi
/ (3 \sqrt{2})$ is the volume fraction of close-packed monodisperse
hard spheres. For instance, for $\Phim = 0.05$ and $\Phif = 0.50$,
crystallites would cause a peak at $r_\mathrm{fcc} \simeq 1.56$ or
$r_\mathrm{bcc} \simeq 1.91$. Clearly, peaks at these positions are
absent in \prettyref{fig:static/gr}. On the other hand, in both
panel (a) and (b) an additional peak at $r \simeq 1.8$ emerges upon
increasing $\Phif$. However, this peak is a well-known feature of
glass-forming systems~\cite{kob1995} and thus provides further
evidence for the absence of crystallization. We conclude that
crystallites, albeit possibly temporarily present, do not play a major
role in the systems under investigation.
\subsection{Kinetic diagrams}%
\label{subsec:Kinetic.diagrams}%
\begin{figure}%
\includegraphics* [width = \Figurewidth]{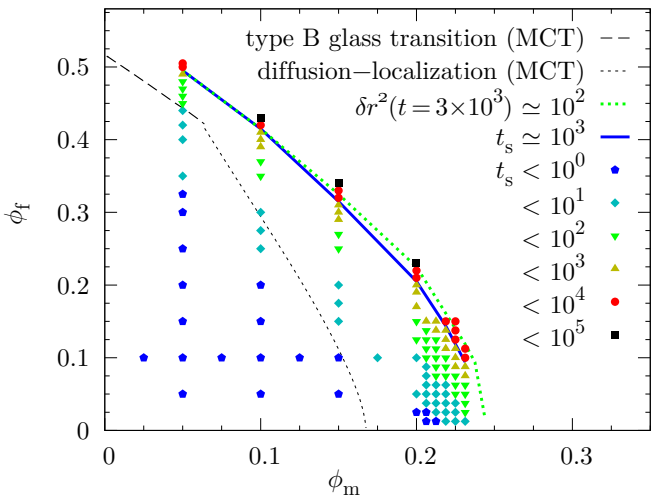}%
\caption{\label{fig:kinetic_diagram/fkt_self}%
(Color online) Kinetic diagram based on single-particle properties
(see text). Symbols: time $\Ts$ needed for $\Fs(k{=}7.0, t)$ to
decay below $\Fs^*=0.1$. Thick solid (blue) line: interpolation
through points for which $\Ts \simeq 10^3$. Thick dotted (green)
line: arrest line based on $\Drs(t)$ from~\cite{kurzidim2009}~(extended to
low $\Phif$ values). Thin dashed and dotted (black)
lines: MCT transition lines pertaining to single-particle
properties from~\cite{krakoviack2009}.}%
\end{figure}%
\begin{figure}%
\includegraphics* [width = \Figurewidth]{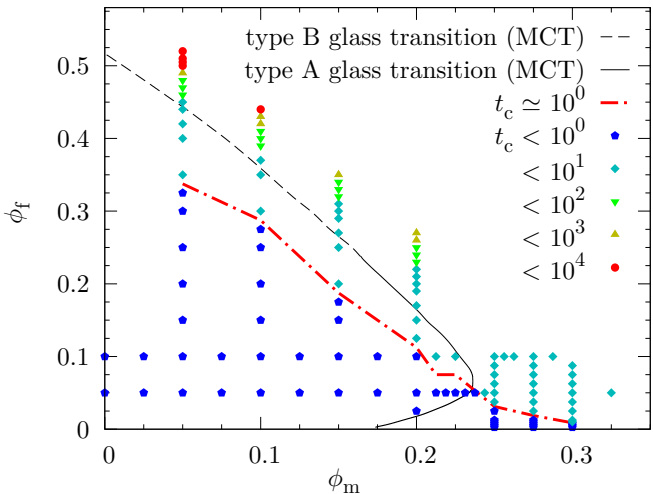}%
\caption{\label{fig:kinetic_diagram/fkt_coll.con}%
(Color online) Kinetic diagram based on collective properties (see
text). Symbols: time $\Tc$ needed for $\Fc(k{=}7.0, t)$ to decay
below $\Fc^*=0.1$. Thick dash-dotted (red) line: interpolation
through points for which $\Tc \simeq 10^0$. Thin dashed and solid
(black) lines: MCT transition lines pertaining to collective
properties from~\cite{krakoviack2009}.}%
\end{figure}%

Many features of the system under investigation can be understood by
means of kinetic diagrams in which the ``state'' of a dynamic property
is indicated in the plane spanned by $\Phim$ and $\Phif$. In an
earlier work~\cite{kurzidim2009} we presented a kinetic diagram
based on the mean-squared displacement~$\Drs(t)$. In order to classify
the state of the system, we chose a mean-squared displacement $\Drs_*
= 100$ and a simulation time~$t_* = 30\,000$; if the system obeyed
$\Drs(t_*) > \Drs_*$ then it was considered ``nonarrested,'' and
otherwise ``arrested.'' The so-constructed kinetic diagram
qualitatively confirmed the behavior of the single-particle dynamics
of the system at hand as predicted by MCT~\cite{kurzidim2009,
krakoviack2009}. The arrest line determined from this criterion is
indicated in \prettyref{fig:kinetic_diagram/fkt_self}.

In \prettyref{fig:kinetic_diagram/fkt_self} we present a kinetic
diagram of the system under investigation based on the
self-intermediate scattering function

\begin{equation} \label{eqn:def_Fskt}%
\Fs(k,t) = \frac
{~~\overline{\Lang \Rtpk(t) \Rtmk(0) \Rang}~~}%
{~~\overline{\Lang \Rtpk(0) \Rtmk(0) \Rang}~~}%
\end{equation}%

\newcommand{\FootDenominator}{\footnote{The denominator, although equal to
unity, has been retained to
unveil the structure of the correlator.}}%

\noindent where $\Rtpk(t) = \exp \{ i \mathbf{k} \cdot \Rt(t) \}$ is
the density of a tagged (superscript ``$\mathrm{t}$'') fluid particle
in $k$ space after some time~$t$ has passed~\FootDenominator. The
symbols in the figure indicate the times $\Ts$ needed for
$\Fs(k{=}7.0, t)$ to decay below the value $\Fs^* = 0.1$. The value $k
= 7.0$ was chosen to be close to the main peak in $S(k)$ in a typical
high-density QA system---for instance, for $\Phim = 0.05$ and $\Phif =
0.50$ (\Cf\ \prettyref{fig:static/sk}) the first peak in $S(k)$ is
located at $k \simeq 7.2$; for $\Phim = 0.25$ and $\Phif = 0.10$ it is
found at $k \simeq 6.6$.

If we let $\Ts^* \simeq 10^3$, discriminating whether for a state
point $\Ts > \Ts^*$ (``arrested'') or $\Ts< \Ts^*$ (``nonarrested'')
yields the thick solid (blue) line in \prettyref{fig:kinetic_diagram/fkt_self}.
Obviously, the latter is only slightly
different from the thick dotted (green) line that is obtained from the
$\Drs(t)$ criterion described above. This is not unexpected since both
$\Fs(k,t)$ and $\Drs(t)$ are single-particle properties, and in the
Gaussian limit it is even $\Fs(k,t) = \exp \{k^2 \Drs(t) / 6\}$~\cite
{hansen1986}. Thus, the validity of the approach chosen in~\cite{kurzidim2009}
is confirmed.

In order to complement the kinetic diagrams based upon single-particle
properties, in \prettyref{fig:kinetic_diagram/fkt_coll.con} we
present for the first time a kinetic diagram for a collective property
of the system. For comparison with the theory developed in~\cite{krakoviack2005,
krakoviack2007, krakoviack2009} we chose to operate
with the connected part of the intermediate scattering function

\begin{equation} \label{eqn:def_Fckt}%
\Fc(k,t) = \frac
{~~\overline{\Lang \Drpk(t) \Drmk(0) \Rang}~~}%
{~~\overline{\Lang \Drpk(0) \Drmk(0) \Rang}~~}%
\end{equation}%

\noindent where $\Drpk(t) = \Rpk(t) - \langle \Rpk(0) \rangle$ is the
fluctuating part of $\Rpk(t) = \sum_i \exp \{ i \mathbf{k} \cdot
\Rf_i(t) \}$ (\Cf\ \prettyref{sec:Discussion}). Very much like for
the self-intermediate scattering function $\Fs(k,t)$, in this figure
we display the times $\Tc$ needed for $\Fc(k{=}7.0, t)$ to decay below
the value $\Fc^* = 0.1$. The thick dash-dotted (red) line in
\prettyref{fig:kinetic_diagram/fkt_coll.con} interpolates through
points for which $\Tc \simeq 10^0$. The shape of this iso-$\Tc$ line
clearly contradicts the MCT scenario~\cite{krakoviack2005,
krakoviack2007}, which predicts a re-entrance regime (a
``bending-back'' of the collective glass transition line) for large
values of $\Phim$.

Since the arrest line obtained from $\Ts$ is situated at considerably
larger $\Phim$ than the diffusion-localization line predicted by MCT,
it is reasonable to expect a similar shift for the glass transition
line. Unfortunately, an estimate for such shift suggests that the
re-entrant scenario may take place at combinations of $\Phim$ and
$\Phif$ at which systems cannot be set up due to geometric constraints
(\Cf\ \prettyref{apx:System.setup.details}). However, one should
also bear in mind that the available MCT calculations are based on
structure factors obtained from integral equations~\cite{krakoviack2005,
krakoviack2007, krakoviack2009} as it is still
difficult to extract reliable blocked correlations from
simulations~\cite{meroni1996, schwanzer2009}. Thus, it remains hard
to disentangle the inherent deficiencies of MCT from those related to
the input structural data, especially at high matrix densities.

Based on simple arguments, in~\cite{krakoviack2009} it is concluded
that the glass transition line has to coincide with the
diffusion-localization line in the limit of vanishing fluid density.
The data presented in \prettyref{fig:kinetic_diagram/fkt_self} and
\ref{fig:kinetic_diagram/fkt_coll.con} are ostensibly inconsistent
with this expectation. We remark, however, that our simulations do not
provide for this special case since we decided to retain a
nonvanishing number of fluid particles in every system realization so
as to always have finite fluid-fluid correlations. More importantly,
in \prettyref{sec:Discussion} we will provide evidence that using
the present definition of the decay times $\Tc$ and $\Ts$, the two
quantities are not quite comparable to one another.
\subsection{Path I: constant low matrix density}%
\label{subsec:Path.I..constant.low.matrix.density}%
\begin{figure}%
\includegraphics* [width = \Figurewidth]{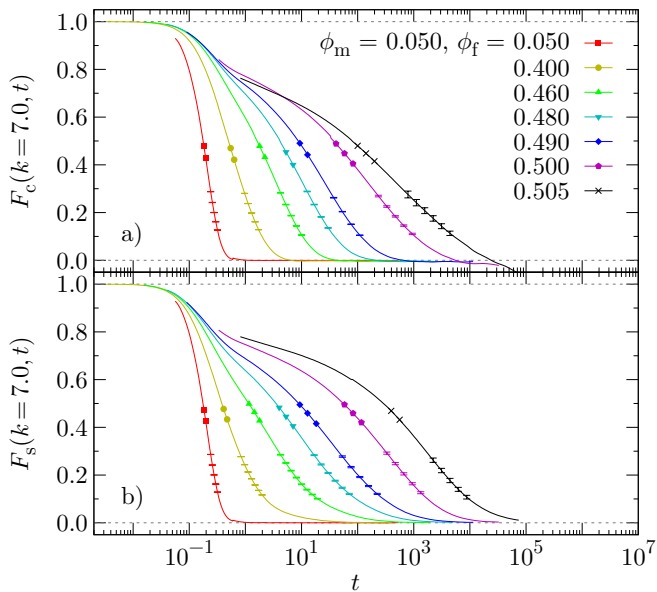}%
\caption{\label{fig:const_mpf0.05/fkt}%
(Color online) Intermediate scattering functions for a series of
$\Phif$ values at fixed $\Phim = 0.05$ and $k=7.0$. (a) Connected
correlator $\Fc(k,t)$, (b) single-particle correlator $\Fs(k,t)$.
Error bars: see \prettyref{fig:static/sk}.}%
\end{figure}%
\begin{figure}%
\includegraphics* [width = \Figurewidth]{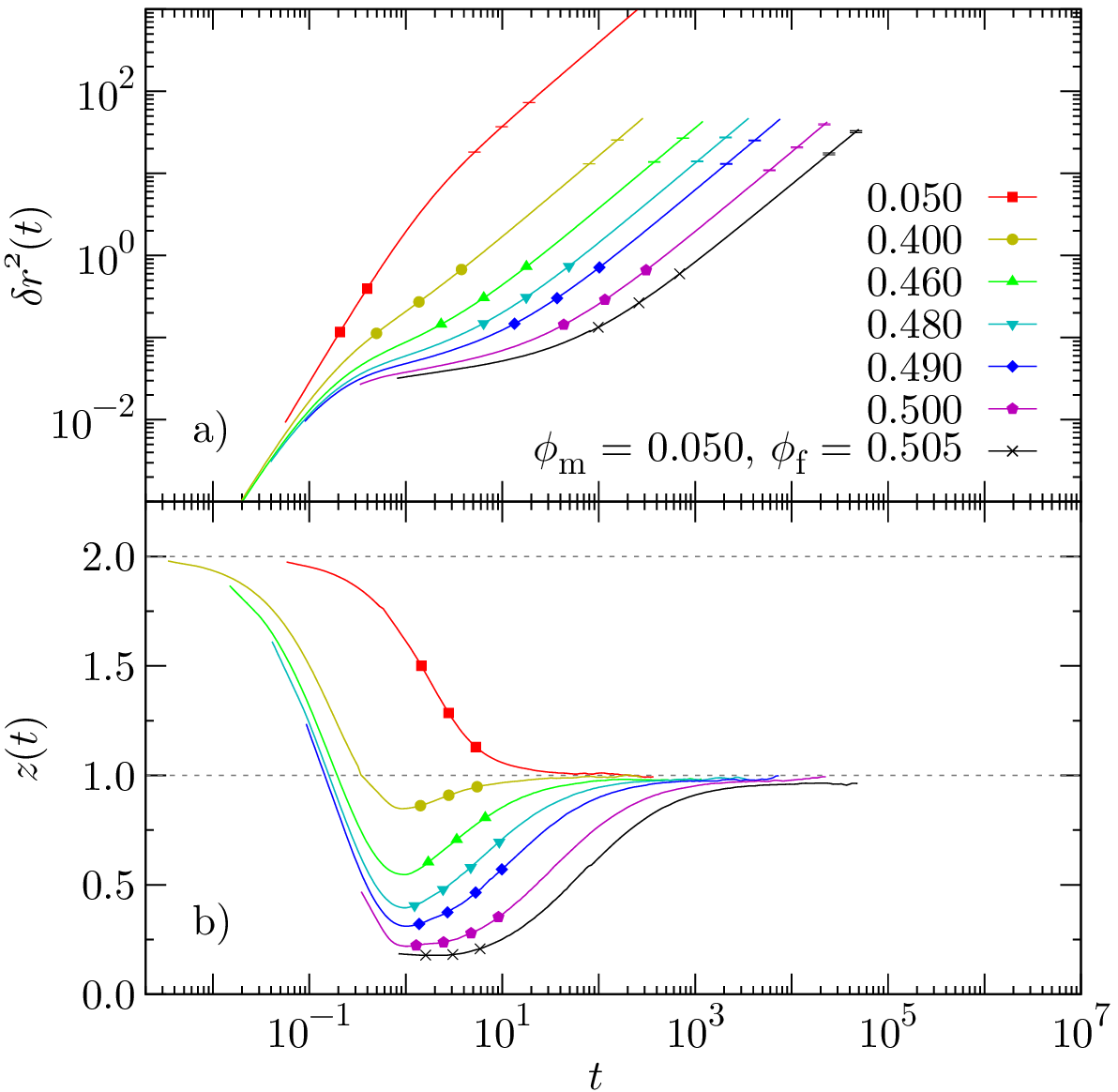}%
\caption{\label{fig:const_mpf0.05/msd}%
(Color online) Mean-squared displacement (a) and its logarithmic
derivative (b) for a series of $\Phif$ values at fixed $\Phim =
0.05$. Error bars [only panel (a)]: see \prettyref{fig:static/sk}.}%
\end{figure}%

In this subsection, as well as in \prettyref{subsec:Path.II..constant.intermediate.matrix.density}
and \ref{subsec:Path.III..constant.fluid.density}, we refine our findings
from~\cite{kurzidim2009} concerning $\Fs(k,t)$, $\Fc(k,t)$, and
$\Drs(t)$. Additionally, we computed the logarithmic derivative

\begin{equation} \label{eqn:def_zt}
z(t) = \frac{d \Lbkt \log \Drs(t) \Rbkt}{d \Lbkt \log t \Rbkt}%
\end{equation}%

\noindent which facilitates the discussion of $\Drs(t)$ since in an
assumed subdiffusive law of the form $\Drs(t) \propto t^z$ it
represents the momentary value of the power $z$. The first set of data
we present for state points at various $\Phif$ at a common $\Phim =
0.05$, which corresponds to a vertical line in the kinetic diagrams
\prettyref{fig:kinetic_diagram/fkt_self} and \ref{fig:kinetic_diagram/fkt_coll.con}
that shall henceforth be referred
to as ``path~I.''

In \prettyref{fig:const_mpf0.05/fkt}, the time dependence of
$\Fc(k,t)$ and $\Fs(k,t)$ is depicted. For both observables, the wave
vector was chosen to be $k = 7.0$, which is close to the first peak in
the static structure factor (\Cf\ \prettyref{fig:static/sk}). For
elevated $\Phif$ an intermediate-time plateau can be identified in
both $\Fs(k,t)$ and $\Fc(k,t)$; as $\Phif$ is increased, this plateau
extends to longer and longer times but does not change in height.
Moreover, $\Fs(k,t)$ and $\Fc(k,t)$ decay on the same time scale as
they approach this type-B transition. This is a well-known phenomenon
in bulk glass formers~\cite{gotze1992}, where the collective glass
transition drives the arrest of the individual particles.

The bulklike behavior is also prevalent in $\Drs(t)$, as is evident
from \prettyref{fig:const_mpf0.05/msd}. For the lowest value of
$\Phif$ depicted, $\Drs(t)$ crosses over directly from ballistic
($z=2$) to diffusive ($z=1$) behavior. Upon increasing $\Phif$ the
ballistic range is followed by a distinct regime in which $\Drs(t)$
grows only very slowly, which is reflected by a drop of $z(t)$ below
unity and as low as $0.15$ for the highest $\Phif$ considered. This is
another manifestation of the cage effect (\Cf\ \prettyref{sec:Introduction}).
For all such $\Phif$, ensuingly diffusive
behavior is recovered in the long-time limit, with the time to
approach $z=1$ increasing enormously with $\Phif$.
\subsection{Path II: constant intermediate matrix density}%
\label{subsec:Path.II..constant.intermediate.matrix.density}%
\begin{figure}%
\includegraphics* [width = \Figurewidth]{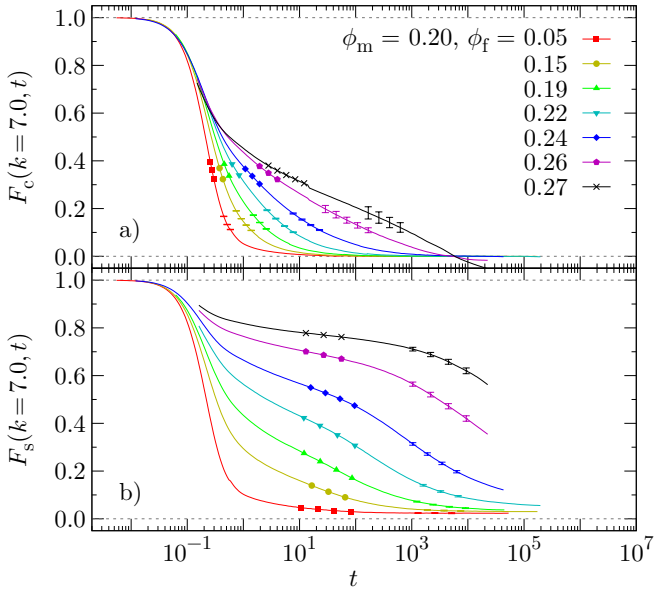}%
\caption{\label{fig:const_mpf0.20/fkt}%
(Color online) Intermediate scattering functions for a series of
$\Phif$ values at fixed $\Phim = 0.20$ and $k=7.0$. (a) Connected
correlator $\Fc(k,t)$, (b) single-particle correlator $\Fs(k,t)$.
Error bars: see \prettyref{fig:static/sk}.}%
\end{figure}%
\begin{figure}%
\includegraphics* [width = \Figurewidth]{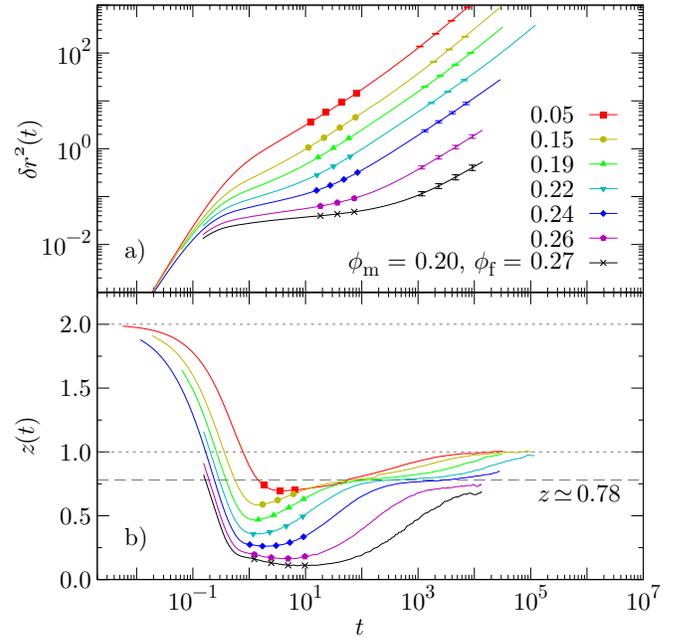}%
\caption{\label{fig:const_mpf0.20/msd}%
(Color online) Mean-squared displacement (a) and its logarithmic
derivative (b) for a series of $\Phif$ values at fixed $\Phim =
0.20$. Error bars [only panel (a)]: see \prettyref{fig:static/sk}.}%
\end{figure}%

In this subsection we present data for state points at various $\Phif$
at a common $\Phim = 0.20$, \Ie, along another vertical line in the
kinetic diagrams presented in \prettyref{subsec:Kinetic.diagrams}.
In the following we will refer to this line as ``path II.''

In \prettyref{fig:const_mpf0.20/fkt}, the time dependence of
$\Fc(k{=}7.0,t)$ and $\Fs(k{=}7.0,t)$ is depicted. As can easily be
seen, the decay pattern of both correlators is quite unlike that in
\prettyref{subsec:Path.I..constant.low.matrix.density} and
\prettyref{fig:const_mpf0.05/fkt} therein. Moreover, $\Fc(k,t)$ and
$\Fs(k,t)$ deviate substantially from one another. Most strikingly,
the time scales on which the correlators decay to their long-time
value are vastly disparate, and increasingly so for larger values of
$\Phif$. At $\Phif = 0.22$, for instance, the decay times read
$\Tc=7.9{\times}10^0$ and $\Ts = 5.4{\times}10^3$. This represents a
difference of almost three decades in time, with the \emph{collective} correlator
decaying faster.

$\Fc(k{=}7.0,t)$ seems to approach a transition since the time it
needs to decay to its long-time value increases substantially as
$\Phif$ is increased. However, it is difficult to pinpoint for sure
whether $\Fc(k,t)$ exhibits a transition of type-A (single-step decay)
or type-B (two-step decay, \Cf\ \prettyref{subsec:Path.I..constant.low.matrix.density}).
For one, in simulations
a type-A transition may show features that differ from the theoretical
predictions~\cite{krakoviack2007}. Further, it is possible that an
intermediate-time plateau would emerge upon increasing $\Phif$ above
$0.27$ (the largest value of $\Phif$ depicted in \prettyref{fig:const_mpf0.20/fkt}).
Note, however, that beyond $\Phif = 0.27$
geometry strongly inhibits to realize system instances (see \prettyref
{apx:System.setup.details}). We conclude that using the current
scheme of simulation and analysis, no definite statement can be made
about the type~of transition in $\Fc(k,t)$ in this part of the kinetic
diagram.

$\Fs(k{=}7.0,t)$ shows an intermediate-time plateau for large values
of $\Phif$, just as in \prettyref{fig:const_mpf0.05/fkt}. The times
at which such plateau may be identified range from $10^0 < t < 10^1$
for $\Phif = 0.15$ up to $10^0 < t < 10^4$ for $\Phif = 0.27$. In
contrast to \prettyref{fig:const_mpf0.05/fkt}, $\Fs(k,t)$ attains
an additional nonzero long-time value that increases with $\Phif$. A
tail can be identified even at the lowest fluid packing fraction
considered ($\Phif = 0.05$), whereas in this case \emph{no}
intermediate-time plateau is present. Therefore, for the sequence of
state points at hand, two superposing decay behaviors exist in
$\Fs(k,t)$: a type-A transition responsible for the long-time value,
and another relaxation mechanism at higher $\Phif$ that leads to the
intermediate-time plateau and is likely to be caused by a weak
collective cage effect. The type-A transition is probably connected to
the continuous diffusion-localization transition predicted by
MCT~\cite{krakoviack2009} and is discussed in more detail in
\prettyref{subsec:Path.III..constant.fluid.density} and \ref{subsec:Trapping}.

Although not as obvious, \prettyref{fig:const_mpf0.20/msd}
evidences that the mean-squared displacement $\Drs(t)$ differs from
bulklike behavior as well. As is the case for $\Phim = 0.05$, for
$\Phim = 0.20$ ballistic behavior is followed by a regime for which
$z$ drops well below unity. Contrary to the bulklike case, this
observation holds also for very low $\Phif$, suggesting that caging by
fluid particles is not the only mechanism effective in this regime.
Most likely, at low $\Phif$ the drop in $z$ is entirely due to the
porous matrix, while at larger $\Phif$ both the pores and caging play
a role.

For low $\Phif$, the range of decreasing $z$ is followed by an almost
straight increase to $z=1$, \Ie, a direct approach of the diffusive
regime. Upon increasing $\Phif$, a distinct intermediate-time plateau
with $z<1$ emerges in $z(t)$, which corresponds to a subdiffusive
regime in $\Drs(t)$. The estimated value $z\simeq 0.78$ in the
subdiffusive regime is only weakly dependent on $\Phif$; merely for
$\Phif \geq 0.26$ the value of $z$ seems to systematically decrease.
This decline, however, may be due to insufficient equilibration of the
respective systems.
\subsection{Path III: constant fluid density}%
\label{subsec:Path.III..constant.fluid.density}%
\begin{figure}%
\includegraphics* [width = \Figurewidth]{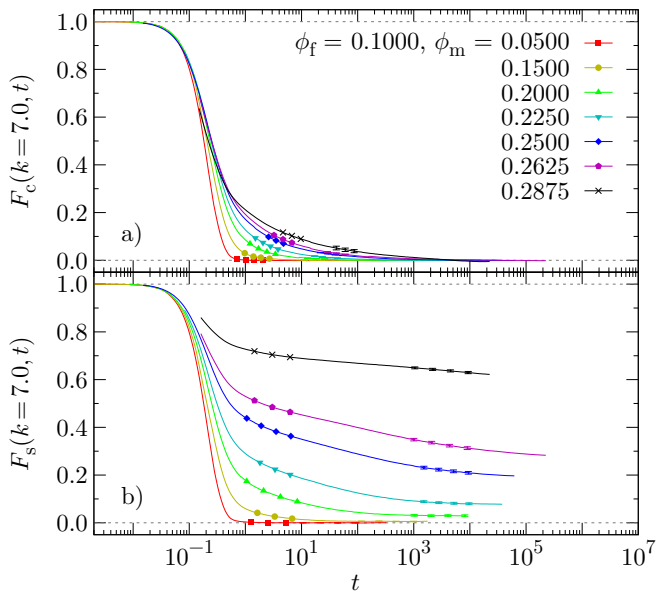}%
\caption{\label{fig:const_fpf0.10/fkt}%
(Color online) Intermediate scattering functions for a series of
$\Phim$ values at fixed $\Phif = 0.10$ and $k=7.0$. (a) Connected
correlator $\Fc(k,t)$, (b) single-particle correlator $\Fs(k,t)$.
Error bars: see \prettyref{fig:static/sk}.}%
\end{figure}%
\begin{figure}%
\includegraphics* [width = \Figurewidth]{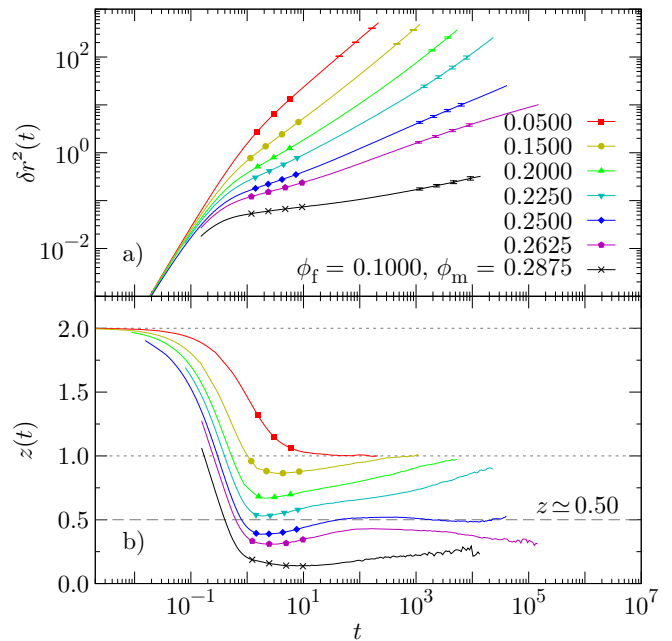}%
\caption{\label{fig:const_fpf0.10/msd}%
(Color online) Mean-squared displacement (a) and its logarithmic
derivative (b) for a series of $\Phim$ values at fixed $\Phif =
0.10$. Error bars [only panel (a)]: see \prettyref{fig:static/sk}.}%
\end{figure}%

Lastly we turn to a selection of state points at varying $\Phim$ for a
relatively low fixed $\Phif = 0.10$, \Ie, along a horizontal straight
line in the kinetic diagrams, perpendicular to paths~I and II. This
line will henceforth be called ``path III.''

In \prettyref{fig:const_fpf0.10/fkt} we present $\Fc(k{=}7.0,t)$
and $\Fs(k{=}7.0,t)$ for these state points. The correlators differ
substantially from each other and from their bulklike
counterparts---even more than was the case in \prettyref{subsec:Path.II..constant.intermediate.matrix.density}.
Most notably,
$\Fs(k,t)$ attains a sizeable nonzero value as time approaches
infinity, and does so more pronouncedly as $\Phim$ increases. By
contrast, for all $\Phim$ represented in \prettyref{fig:const_fpf0.10/fkt}
$\Fc(k,t)$ decays strictly to zero. On the
other hand the actual pattern of the decay to the long-time value is
quite similar in $\Fc(k,t)$ and $\Fs(k,t)$: unlike in \prettyref{fig:const_mpf0.20/fkt},
both correlators decay in a single step,
suggesting only type-A transitions to be involved.

Comparison of \prettyref{fig:const_fpf0.10/msd} with \prettyref{fig:const_mpf0.20/msd}
and \ref{fig:const_mpf0.05/msd} reveals the
functional form of $\Drs(t)$ to be more sensitive to $\Phim$ than to
$\Phif$. As observed in \prettyref{subsec:Path.I..constant.low.matrix.density},
at low matrix densities
the initial ballistic behavior is directly followed by diffusive
behavior. For intermediate $\Phim< 0.25$ (\Cf\ \prettyref{subsec:Path.II..constant.intermediate.matrix.density})
a region of
very slowly increasing particle displacement ($z<1$) succeeds the
ballistic range. The value of the exponent $z$ in this region
decreases as $\Phim$ increases. Subsequently, diffusive behavior is
recovered as $z$ raises in an almost linear fashion.

Upon further increasing $\Phim$, at some $\Phim^*$ a geometric
percolation transition takes place in the space accessible to the
fluid particles (``voids''), with a void stretching through the whole
system at $\Phim \leq \Phim^*$. This transition is intimately
connected with the diffusion-localization transition predicted by
MCT~\cite{krakoviack2009} and observed in our simulations. As a
hallmark of this transition, $\Drs(t)$ does not approach diffusive
behavior for $\Phim \simeq 0.25$ but instead remains at an
approximately constant $z \simeq 0.5$ for a window covering about
three decades in time. Since for that matrix density $z(t)$ ultimately
increases beyond $0.5$, one can expect the diffusion-localization
transition to take place at slightly higher $\Phim$. An upper bound to
that value is $\Phim = 0.2625$, the smallest value presented that is
larger than $0.25$. At this packing fraction of obstacles the window
of constant $z$ extends over roughly two decades in time, with $z$
ultimately decreasing below that plateau.

Judging from these observations, the subdiffusion exponent~$z$ at the
diffusion-localization transition might be slightly lower than 0.5.
Nevertheless, the observed value of $z$ is in striking agreement with
MCT, which predicts that $z = 0.5$~\cite{krakoviack2009}. An
additional agreement of numbers calls for investigation: It has long
been known in theory~\cite{kaerger1992} and also verified in
experiment~\cite{wei2000} that in a one-dimensional random
walk---so-called ``single-file diffusion''---the mean-squared
displacement of the particles grows with precisely the same exponent
as in our case. However, it remains an open question whether in QA
systems the main effect in single-file diffusion---nonovertaking
particles---is prevalent, or if other (possibly compensating) reasons
lead to a coincidental agreement in $z$.
\subsection{Trapping}%
\label{subsec:Trapping}%
\begin{figure}%
\includegraphics* [width = \Figurewidth]{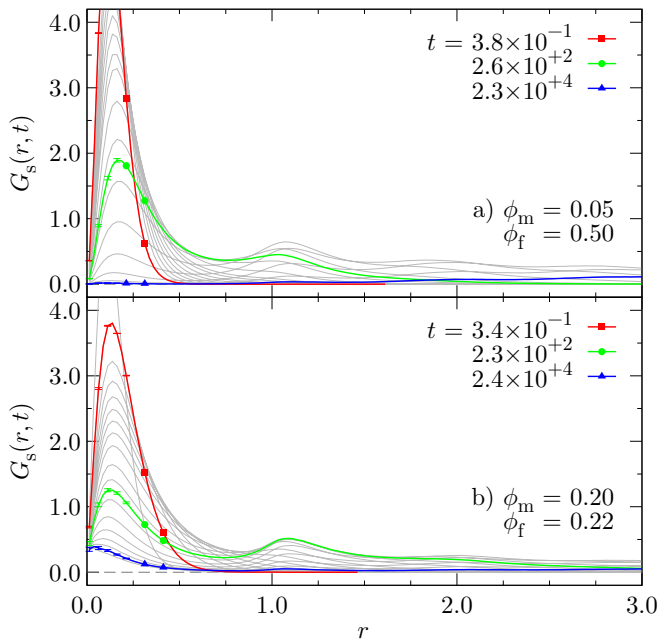}%
\caption{\label{fig:trapping/upper-left}%
(Color online) self-part of the direction-averaged van Hove
function $\Gs(r,t)$ for (a) $\Phim = 0.05$, $\Phif = 0.50$ (b)
$\Phim = 0.20$, $\Phif = 0.22$. Curves are approximately
logarithmically spaced in time; some are highlighted for visual
guidance. Error bars: see \prettyref{fig:static/sk}.}%
\end{figure}%
\begin{figure}%
\includegraphics* [width = \Figurewidth]{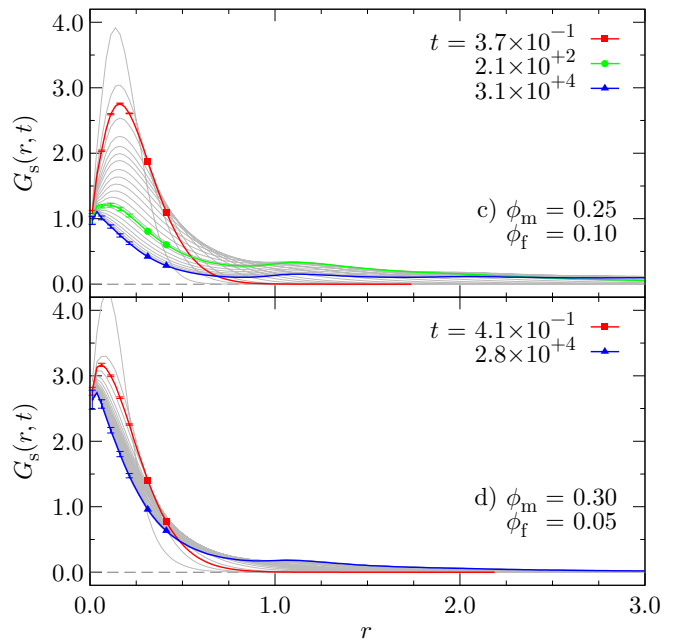}%
\caption{\label{fig:trapping/lower-right}%
(Color online) self-part of the direction-averaged van Hove
function $\Gs(r,t)$ for (c) $\Phim = 0.25$, $\Phif = 0.10$ (d)
$\Phim = 0.30$, $\Phif = 0.05$. Curves are approximately
logarithmically spaced in time; some are highlighted for visual
guidance. Error bars: see \prettyref{fig:static/sk}.}%
\end{figure}%

In this subsection we will elucidate some aspects of the geometrical
structure imposed by the quenched matrix upon the fluid immersed
therein. The key concept involved is that of ``voids,'' \Ie, domains
of space that may be fully explored by a fluid particle if placed
within. In HS-QA systems distinct voids are separated by infinite
potential walls, and there exist voids of finite volume. Particles
located in such finite void cannot travel infinitely far away from
their initial location; such particles will henceforth be denoted as
``trapped particles,'' and the corresponding void as a ``trap.'' Due
to the statistical nature of the matrix structure, at any nonzero
matrix density there exist traps. Likewise, for any nonzero density of
traps there will be trapped particles since the initial locations of
the fluid particles are randomly distributed.

Two types of nontrivial questions remain to be assessed. First of all,
there exists the aforementioned distinct $\Phim^*$ above which all
fluid particles are trapped. Differently phrased (\Cf\ \prettyref{subsec:Path.III..constant.fluid.density}),
upon varying $\Phim$, a
percolation transition of the space accessible to the fluid particles
takes place: for $\Phim \leq \Phim^*$, there exists an
infinitely-large void, whereas for $\Phim > \Phim^*$ any void is of
finite size. Naturally, the question arises what the precise value of
$\Phim^*$ is. In the following we will investigate the behavior of
fluid particles that explore the voids by studying the distribution of
their displacements, as was done before for other types of
confinement~\cite{hofling2006}.

Our quantity of choice for the desired analysis is the self-part of
the direction-averaged van Hove correlation function

\begin{equation} \label{eqn:def_Gsrt}%
\Gs(r,t) = \overline{\Lang
\delta \Lpar r - \Delta r(t) \Rpar
\Rang}%
\end{equation}%

\noindent where $\Delta r(t) = |\Rt(t)-\Rt(0)|$ and by construction
$\int_0^\infty \Gs(r,t)\,dr = 1$. Strictly speaking, to assess the
void structure only knowledge about the infinite-time limit
$\Gs(r,t{\rightarrow}\infty)$ is required. However, since data are
readily available, in the following we will also discuss some features
of the time evolution of $\Gs(r,t)$ in order to corroborate and extend
our findings from the previous sections.

Using the self-part of the van Hove function a number of quantities of
interest can be assessed at least qualitatively. For instance, the
matrix density at percolation, $\Phim^*$, can be estimated by
exploiting the fact that as the system control parameter (here
$\Phim$) is varied the average size of the ``aggregates'' (in this
case the traps) diverges at the transition. For sufficiently small
$r$, the distribution of trap sizes is reflected in
$\Gs(r,t{\rightarrow}\infty)$ since in the infinite-time limit the
nontrapped fluid particles have diffused far away. Concentrating on a
particular distance $\tilde r \sim \sigma$, the function $\Gs(\tilde
r,t{\rightarrow}\infty; \Phim)$ will exhibit two maxima as $\Phim$ is
varied, one for $\Phim \lesssim \Phim^*$ and one for $\Phim \gtrsim
\Phim^*$. The corresponding integral $\int_{0}^{\tilde r}%
\Gs(r,t{\rightarrow}\infty)\,dr$ represents an estimate for the
fraction of fluid particles located in traps. Of course, in
simulations $t{\rightarrow}\infty$ has to be approximated by some
finite time $\tilde t$ greater than the structural relaxation time of
the system.

In \prettyref{fig:trapping/upper-left} and \ref{fig:trapping/lower-right}
the van Hove function is displayed at
various $t$ as a function of $r$. For visual guidance, curves are
highlighted for selected times. The times of successive light (gray)
lines differ by ${\sim}\sqrt[6\,]{10}$, \Ie, there are about six
curves per time decade. In \prettyref{fig:trapping/upper-left}(a)
$\Gs(r,t)$ is shown for $\Phim = 0.05$ and $\Phif = 0.50$, a bulklike
state point like those discussed in \prettyref{subsec:Path.I..constant.low.matrix.density}.
At short times
$\Gs(r,t)$ exhibits a single, approximately Gaussian peak. As $t$
increases, the maximum value of this peak decreases, but notably its
$r$ position remains nearly constant even for large $t$. This is
another manifestation of the cage effect prevalent in glassy
systems~\cite{kob1995}. Further support for this picture arises
from the second peak in $\Gs(r,t)$ that emerges at $r \simeq 1.1$ for
$t \gtrsim 10^2$, which indicates the presence of hopping processes.
For very long times even a moderate third peak appears at $r \simeq 2$
as $\Gs(r,t)$ recovers a Gaussian-like shape centered at $r > 3$.

The validity of these arguments becomes clear when considering
\prettyref{fig:trapping/upper-left}(b) which depicts $\Gs(r,t)$ for
$\Phim = 0.20$ and $\Phif = 0.22$. For this state point, representing
intermediate matrix density, similar features can be observed for
short and intermediate times, only with the first peak decreasing
slower in height, the second peak being less pronounced, and the third
peak missing. However, for large $t$ and small $r$ the shape of
$\Gs(r,t)$ is markedly different from that in \prettyref{fig:trapping/upper-left}(a):
even for $\tilde t = 7.1{\times}10^4$
(the largest time considered) we find $\Gs(r,\tilde t)$ to be nonzero
for small $r$. Since curves of $\Gs(r,t)$ very much resemble each
other for $t \lesssim \tilde t$ and $r <0.5$, it is reasonable to
assume that in this range $\Gs(r,\tilde t)$ represents a reliable
approximation to $\Gs(r,t{\rightarrow}\infty)$. Hence we can infer the
presence of trapped particles, with their fraction being $x \simeq
\int_0^{\tilde r} \Gs(r,\tilde t)\,dr = 8\%$ using $\tilde r = 0.5$.
Moreover, since $\Gs(r,\tilde t)$ is close to zero for $r> \tilde r$,
there likely exist only few traps with a spatial extent exceeding
$\tilde r$.

The trend seen from \prettyref{fig:trapping/upper-left}(a) to \ref{fig:trapping/upper-left}(b)
is continued by the systems represented
in \prettyref{fig:trapping/lower-right}. Panel (c) of that figure
shows the state point at $\Phim = 0.25$ and $\Phif = 0.10$, that is,
close to the diffusion-localization transition (discussed in
\prettyref{subsec:Path.III..constant.fluid.density}); the state
point in panel (d) is situated well in the localized phase at $\Phim =
0.30$ and $\Phif = 0.05$. The change in shape over time becomes less
pronounced as $\Phim$ increases, turning the second peak into a mere
shoulder.

The largest times depicted in \prettyref{fig:trapping/lower-right}
again represent a sound approximation of
$\Gs(r,t{\rightarrow}\infty)$. For the state point in panel (d) we
numerically obtain $x \simeq \int_0^{\tilde r} \Gs(r,\tilde t)\,dr =
98\%$ using $\tilde r = 3$ and $\tilde t = 7.8{\times}10^4$, which is
in excellent agreement with the figure $x = 100\%$ that would be
expected in an infinitely-large system residing beyond the percolation
transition. For the state point in (c), using the same $\tilde r$ and
$\tilde t$ we obtain $x \approx 58\%$, which, however, is rather
imprecise since $\Gs(\tilde r, \tilde t)$ is considerably greater than
zero.

In \prettyref{fig:trapping/lower-right}(c) and (d) the correlator
$\Gs(r,t)$ differs markedly from that in \prettyref{fig:trapping/upper-left}(a)
and (b). For one, in both (c) and (d) a
second peak is present even for $t \rightarrow \infty$, suggesting
that on average traps are large enough to allow for hopping. Another
feature is less obvious but more important to the percolation
transition: for large distances $\Gs(r,t{\rightarrow}\infty)$ is \emph
{greater} in (c) than in (d) for the same $r$. This implies that the
mean trap size is larger in (c) than in (d), which is not
unexpected---in fact this confirms the transition to take place
between (b) and (d). Thus, using $\Gs(r,t)$ we find $0.22 < \Phim^* <
0.30$ as an estimate for the matrix packing fraction at the
percolation transition, which is in accordance with our findings in
the previous sections.
\section{Discussion}%
\label{sec:Discussion}%
\begin{figure}%
\includegraphics* [width = \Figurewidth]{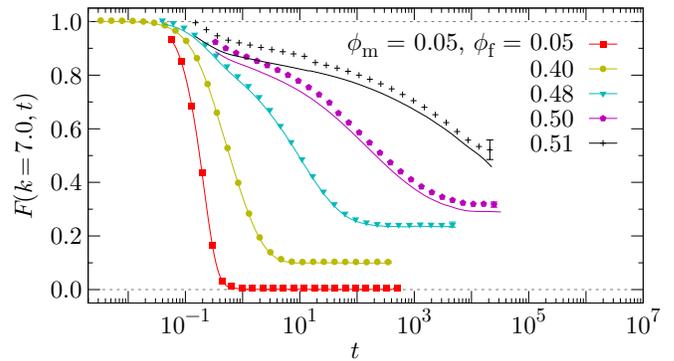}%
\caption{\label{fig:const_mpf0.05/plateau}%
(Color online) Total intermediate scattering function $F(k,t)$
(lines) for selected $\Phif$ values at $\Phim = 0.05$ and $k =
7.0$. Symbols denote the right-hand side of \prettyref{eqn:chk_Fkt}. Error
bars: see \prettyref{fig:static/sk}.}%
\end{figure}%
\begin{figure}%
\includegraphics* [width = \Figurewidth]{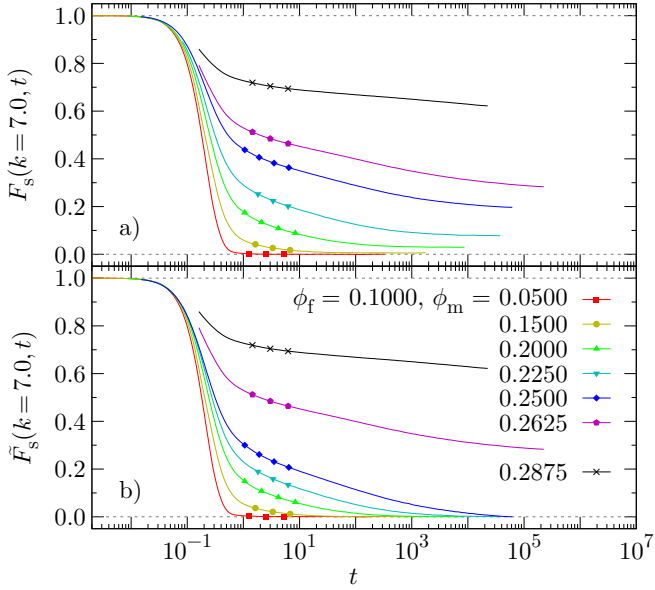}%
\caption{\label{fig:const_fpf0.10/plateau}%
(Color online) self-intermediate scattering functions for a series
of $\Phim$ values at fixed $\Phif = 0.10$ and $k = 7.0$. (a)
self-correlator $\Fs(k,t)$, (b) self-correlator without long-time
plateau $\Fts(k,t)$. For the two largest values of $\Phim$ in
panel (b) it was not possible to reliably identify the long-time
plateau. Therefore the corresponding self-correlators are left
unshifted.}%
\end{figure}%
\begin{figure}%
\includegraphics* [width = \Figurewidth]{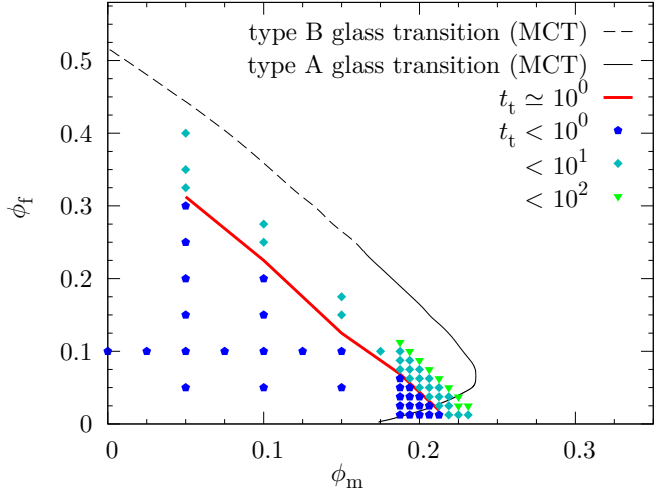}%
\caption{\label{fig:kinetic_diagram/fkt_coll.tot}%
(Color online) Kinetic diagram based on $F(k,t)$. Symbols: time
$\Tt$ needed for $F(k{=}7.0, t)$ to decay below $F^* = 0.1$. Thick
solid (red) line: interpolation through points for which $\Tt
\simeq 10^0$.}%
\end{figure}%
\begin{figure}%
\includegraphics* [width = \Figurewidth]{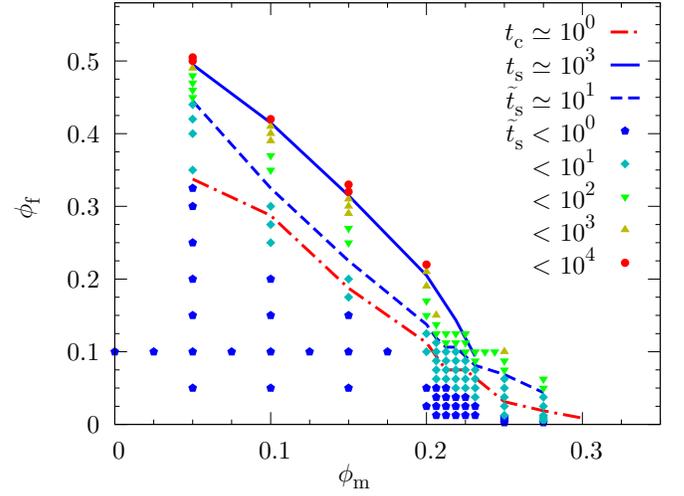}%
\caption{\label{fig:kinetic_diagram/fkt_self.no_ltv}%
(Color online) Kinetic diagram based on $\Fts(k,t)$. Symbols: time
$\Tts$ needed for $\Fts(k{=}7.0, t)$ to decay below $\Fts^* =
0.1$. Dashed (blue) line: interpolation through points for which
$\Tts \simeq 10^1$. Solid (blue) line: interpolation through
points for which $\Ts \simeq 10^3$ (from \prettyref{fig:kinetic_diagram/fkt_self}).
Dash-dotted (red) line:
interpolation through points for which $\Tc \simeq 10^0$ (from
\prettyref{fig:kinetic_diagram/fkt_coll.con}).}%
\end{figure}%

In our presentation of the global aspects of the system's dynamics, as
well as of its behavior along selected paths in the $\{\Phim, \Phif\}$
plane, we have focused our attention on self and connected density
correlators as indicators of the single-particle and collective
dynamics. This choice is motivated by the key role played by these
dynamic correlation functions within the MCT framework for confined
fluids, which provides detailed predictions for $\Fs(k,t)$ and
$\Fc(k,t)$. Before discussing in more detail the comparison between
the theoretical scenario of MCT and our simulations, it is instructive
to study correlation functions \emph{related} to $\Fs(k,t)$ and
$\Fc(k,t)$. Namely, we will investigate here the total intermediate
scattering function

\begin{equation} \label{eqn:def_Fkt}
F(k,t) = \frac{\overline{\Lang \Rpk(t)\Rmk(0) \Rang}}{S(k)}%
\end{equation}%

\noindent as well as a modification of the self-intermediate
scattering function in which the long-time plateau (as observable, for
instance, along paths II and III) is subtracted out. This will allow
to clarify aspects related to the kinetic diagram of the system and to
facilitate the interpretation of the numerical results in the light of
the MCT predictions.

By construction, fluids confined in porous media possess nonzero
average density fluctuations, $\langle \Rpk \rangle \neq 0$, induced
by the matrix structure. This purely-static background becomes visible
in the long-time limit of the total intermediate scattering function
$F(k,t)$. Collective relaxation phenomena in confined fluids are
therefore more conveniently described by the connected correlator,
\prettyref{eqn:def_Fckt}, in which only the \emph{fluctuations}
of the microscopic density are considered. Nevertheless it is
interesting to also inspect the shape of $F(k,t)$ as a function of the
state parameters. In \prettyref{fig:const_mpf0.05/plateau} we show
$F(k,t)$ for some selected densities along path I ($\Phim = 0.05$) for
the same value of $k$ considered in \prettyref{fig:const_mpf0.05/fkt}. Except
for the lowest fluid density, $\Phif = 0.05$, the $F(k,t)$
correlator attains a finite, sizeable plateau at long times. To check
whether this long-time plateau is due to the purely-static background
mentioned above or actually reflects a nonergodic behavior of the
system, we also include in this figure the function $[\Sc(k)\Fc(k,t) +
\Sb(k)]/S(k)$, where

\begin{equation} \label{eqn:def_Sbk}%
\Sb(k) = \overline{\Lang \Rpk \Rang \Lang \Rmk \Rang}%
\end{equation}%

\noindent is the so-called blocked structure factor, and $\Sc(k) =
S(k) - \Sb(k)$ is the connected structure factor. It is easy to show
that, if the system is ergodic, the blocked structure factor is the
infinite-time limit of the un-normalized total intermediate scattering
function~\cite{krakoviack2007} and that

\begin{equation} \label{eqn:chk_Fkt}
F(k,t) = \frac{\Sc(k)\Fc(k,t) + \Sb(k)}{S(k)}%
\end{equation}%

\noindent holds as an equality.

From \prettyref{fig:const_mpf0.05/plateau} we see that this
equality is indeed fulfilled for $\Phif \alt 0.48$, while slight
deviations occur at the largest fluid packing fractions. Note,
however, that the calculation of $\Sb(k)$ is also affected by
statistical noise (see error bars in \prettyref{fig:const_mpf0.05/plateau})
and systematic effects~\cite{schwanzer2009}. Nonetheless \prettyref{eqn:chk_Fkt}
provides an
effective criterion to discriminate nonequilibrated samples as far as
the collective relaxation is concerned. We also note that if we employ
the criterion $F(k{=}7.0, \Tt) = 0.1$ to define the relaxation time
$\Tt$, the latter strongly overestimates the slowness of the
collective relaxations compared to $\Tc$ due to the emergence of the
finite plateau in $F(k,t)$ upon increasing $\Phif$. This should be
born in mind when discussing decoupling phenomena between
single-particle and collective properties.

\newcommand{\FootSubPlateau}{\footnote{The numerical procedure to evaluate
$\Fts(k,t)$ for fixed $k$ is as
follows: first, the global minimum $M = \min_{t > 0} \Fs(k,t)$ of
the self-intermediate scattering function is identified. Then, the
standard deviation, $S$, of $\Fs(k,t)$ is computed over different
time ranges $\{ t_\mathrm{min}, t_\mathrm{max} \}$, keeping the
upper bound $t_\mathrm{max}$ of the range fixed at the maximum time
available while progressively increasing the lower bound
$t_\mathrm{min}$. If $S/(1-M)$ is below some preset tolerance
threshold then $\Fs(k,t)$ is considered to be constant in the
corresponding time range. If this condition is met while
$t_\mathrm{max} / t_\mathrm{min} > 10$ then $\Fts(k,t) =
(\Fs(k,t)-M)/(1-M)$. Otherwise, $\Fs(k,t)$ has not reached its
long-time plateau within the time range considered.}}%

A similar, yet physically different long-time plateau is observed in
the self-intermediate scattering functions along path II and III (see
\prettyref{fig:const_mpf0.20/fkt} and \ref{fig:const_fpf0.10/fkt}). As discussed
in \prettyref{subsec:Trapping}, the finite value
of $\Fs(k, t{\rightarrow}\infty)$ reflects the trapping of particles
in disconnected voids of the matrix structure. Since the existence of
a rising plateau affects the evaluation of the relaxation times $\Ts$,
it is sensible to define a modified correlator $\Fts(k,t) = \Fs(k,t) -
\Fs(k, t{\rightarrow}\infty)$, where the long-time limit $\Fs(k,
t{\rightarrow}\infty)$ is identified by some
construction~\FootSubPlateau. The results of this procedure are
collected in \prettyref{fig:const_fpf0.10/plateau} for selected
state points at $\Phif = 0.10$ (path III) and compared to the
corresponding $\Fs(k,t)$. At the two largest $\Phim$ the relaxation to
the finite plateau is not complete within the time range considered,
and the correlators are left unshifted. Evaluation of the
single-particle relaxation time $\Tts$, defined by $\Fts(k{=}7.0,
\Tts) = 0.1$ as for the other correlators, allows to filter out the
contribution to the single-particle dynamics due to trapping.

We constructed two alternative kinetic diagrams for the system at hand
using the relaxation times $\Tt$ and $\Tts$ obtained from $F(k,t)$ and
$\Fts(k,t)$. As in \prettyref{subsec:Kinetic.diagrams}, lines drawn
in \prettyref{fig:kinetic_diagram/fkt_coll.tot} and \ref{fig:kinetic_diagram/fkt_self.no_ltv}
correspond to a fixed, constant
value of the relaxation times (isorelaxation times lines). Let us
first focus on \prettyref{fig:kinetic_diagram/fkt_coll.tot},
displaying the kinetic diagram based on $\Tt$. In this case, the range
of state points available to determine the isorelaxation times lines
is more limited due to the incipient growth of the plateau in the
region of large $\Phif$ and large $\Phim$. Nevertheless it is clear
that the shape of the estimated arrest line is non-re-entrant,
confirming the analysis based on the connected correlators (see
\prettyref{fig:kinetic_diagram/fkt_coll.con}).

A more interesting effect is observed in \prettyref{fig:kinetic_diagram/fkt_self.no_ltv}
where we show the kinetic
diagram based on $\Tts$. For better comparison with the analysis in
\prettyref{subsec:Kinetic.diagrams} we also include the
isorelaxation times lines obtained from the self and the connected
correlators. Interestingly, the shape of the kinetic diagram for
$\Tts$ strongly resembles that obtained from the connected correlator.
In particular, the iso-$\Tts$ lines do not bend rapidly toward $\Phif
= 0$ as $\Phim$ is increased but rather stretch to larger $\Phim$
values, following the trend of the iso-$\Tc$ line. This effect can be
understood as follows: after subtracting the long-time plateau, which
is due to trapping, the residual relaxation of $\Fs(k,t)$ is governed
by a weak, collective caging effect. The relaxation of $\Fs(k,t)$
\emph{toward the long-time limit} occurs, in fact, on a \emph{similar} time
scale as the relaxation of $\Fc(k,t)$. This supports
our interpretation of the complex relaxation pattern of $\Fs(k,t)$ for
large matrix packing fractions (see \prettyref{fig:const_mpf0.20/fkt}) as
a superposition of trapping and caging effects, and confirms the
connection between the diffusion-localization transition and the
appearance of a nontrivial decoupling between single-particle and
collective properties~\cite{kurzidim2009}.
\section{Conclusion}%
\label{sec:Conclusion}%
In this contribution we have investigated
the dynamic properties of a
fluid confined in a disordered porous matrix. In our investigations we
have reduced such system to a simple QA model where the matrix is a
quenched configuration of a liquid and the fluid particles equilibrate
within that matrix. Both matrix and fluid particles are chosen to be
hard spheres of equal size. Thus, the parameter space characterizing
the system reduces to a two-dimensional plane spanned by the packing
fraction of the matrix particles and that of the fluid particles.
Investigations have been carried out via event-driven molecular
dynamics simulations, observables were evaluated using the
double-averaging recipe characteristic for QA systems. Specifically,
we have evaluated the connected and the single-particle intermediate
scattering functions, the mean-squared displacement, and the self-part
of the van Hove function.

The data and conclusions summarized here represent a counterpart to
the recently presented results obtained from a theoretical
framework~\cite{krakoviack2005, krakoviack2007, krakoviack2009}.
The latter approach---an extension of MCT to the QA protocol---had
predicted a number of intriguing features in the kinetic diagram. Our
investigations have complemented this scenario of \textit{ideal}
transitions, unveiling the properties of the \textit{real}
transitions: (i) At low matrix packing fractions we have undoubtedly
confirmed the predicted type-B glass transition scenario. (ii) For
intermediate matrix packing fractions we have found at least two
arrest mechanisms of different nature, with the self-intermediate
scattering function displaying a more complex behavior than its
connected counterpart. This scenario can be rationalized as a
superpositioning of the effect of trapping in disconnected voids and
of a weak caging mechanism collective in nature. (iii) Varying at low
fluid packing fractions the density of the matrix we found evidence
for a diffusion-localization transition, reflected in the
self-intermediate scattering function as well as in the mean-squared
displacement and in the self-part of the van Hove function. (iv)
Despite considerable effort we were unable to support the predicted
re-entrant transition scenario.

In order to obtain deeper insights into the open questions we have
additionally considered along path II the total intermediate
scattering function, and along path III we have subtracted from the
self-intermediate scattering functions the long-time plateau as there
is evidence that the latter is due to trapping. Our analysis of these
data supports the previously established notion of an interplay
between the two phenomena dominating the arrest transition at
intermediate matrix packing fractions, namely trapping and caging.
Work to disentangle the effects of the latter on the dynamic
correlation functions is currently underway.

\begin{acknowledgments}%

The authors gratefully acknowledge helpful discussions with V.
Krakoviack and K. Miyazaki. This work was funded by the Austrian
Science Foundation (FWF) under Project No. P19890-N16.

\end{acknowledgments}
\appendix
\section{System setup details}%
\label{apx:System.setup.details}%
\begin{figure}%
\includegraphics* [width = \Figurewidth]{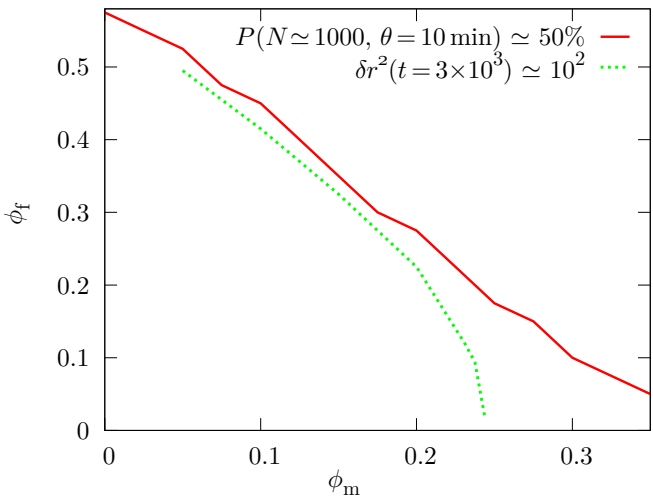}%
\caption{\label{fig:kinetic_diagram/inflation}%
(Color online) Solid (red) line: probability $P(N,\theta) = 50\%$
to set up a system instance with $N = \Nm{+}\Nf \simeq 1000$
particles within $\theta = 10$ CPU minutes using the custom
algorithm described in \prettyref{apx:System.setup.details}.
Lower left: $P(N,\theta) > 50\%$, upper right: $P(N,\theta) <50\%$. Dotted
(green) line: $\Drs(t)$ dynamic arrest criterion
(see \prettyref{subsec:Kinetic.diagrams} and~\cite{kurzidim2009}).}%
\end{figure}%

As mentioned in \prettyref{sec:Model.and.Methods} the initial setup
of a HS-QA system represents a formidable challenge. In this appendix
we will point out algorithms known to succeed in setting up disordered
one-component HS systems at high density, comment on problems in
extending those algorithms to the QA protocol, and report on our
custom solution to the problem.

The challenge is the following: Given the positions of the $\Nm$
immobile obstacle particles, permissible positions are sought for the
$\Nf$ particles that during MD will be allowed to move. Unfortunately,
to this respect multiple complications arise in HS-QA systems, the
three most important of which are: (1) particle overlaps are strictly
forbidden, (2) in the resulting configuration the matrix particles
have to occupy precisely-specified locations, and (3) all
voids---notably also the disconnected ones---have to be considered as
locations for the fluid particles.

The simplest method---trial-and-error insertion of the fluid
particles---is useless already for setting up \emph{bulk}
high-density systems of hard spheres. For the latter task a useful
straight-forward method (although prone to crystallization) is
compressing a low-density system, for instance by coordinate
re-scaling combined with random particle displacements, or by applying
a unidirectional force in a bounded system. However, there also exist
elaborate algorithms to maximize the density in disordered HS systems;
they employ for instance serial deposition~\cite{bennett1972} or
overlap elimination~\cite{jodrey1985}.

Unfortunately, the complications mentioned before render these methods
difficult to adapt and/or ineffective. Therefore, in order to
efficiently achieve high densities in HS-QA systems we devised a
custom algorithm, which consists of the following three steps:

(1) Upon insertion of a fluid particle its ``real'' diameter (the
diameter to be used during MD simulations) is reduced to some minute
value (``deflated'') so that the probability of an insertion via
trial-and-error is greatly increased. This way, throughout the search
of an allowed configuration all particles can be present in the system
simultaneously.

(2) In order to find such allowed configuration a simple Metropolis
Monte-Carlo algorithm~\cite{metropolis1953} is employed using the
HS potential. Since using this algorithm with the minute diameters is
meaningless, at the beginning of each sweep (one displacement attempt
for each fluid particle) the diameter of each deflated particle is
increased (``inflated'') to the maximum value possible without
overlapping another particle. If for a particle the latter value is
greater than its real diameter then inflating is done for this
particular particle and it is assigned its real diameter (``fully
inflated'').

(3) A disconnected void may be filled with (almost) any number of
deflated particles; however, the particular void may be too small to
accommodate all of these particles if they were fully inflated. To
remedy this problem, before certain sweeps all particles that are not
yet fully inflated are removed from the simulation box and then
re-deflated and re-inserted according to step~1. The number of sweeps
separating two such removal-and-reinsertion procedures is gradually
increased so as to allow for any number of sweeps to fully inflate a
particle while still quickly ``draining'' crammed voids.

Obviously, as soon as all particles are fully inflated the setup
routine is completed and a configuration fulfilling the QA
requirements has been found. In step 2 the fluid particles are
inflated serially. This leads to the distribution of diameters
covering a wide range at all times during iteration, except when close
to the approach of an allowed configuration. Since this introduces a
``fluctuating polydispersity,'' we found the generation of ordered
states at high densities to be strongly suppressed even for bulk
monodisperse systems.

Figure \ref{fig:kinetic_diagram/inflation} gives an overview over
the state points at which, according to the above-described steps, a
system instance with $N = \Nm{+}\Nf \simeq 1000$ particles may be set
up with probability $P(N,\theta) \simeq 50\%$ within $\theta = 10$ CPU
minutes (Intel Core 2 Duo E8400). Comparing the so-found line of
discrimination (red solid line in \prettyref{fig:kinetic_diagram/inflation})
with the $\Drs(t)$ dynamic arrest
criterion (green dotted line in \prettyref{fig:kinetic_diagram/inflation};
see \prettyref{subsec:Kinetic.diagrams} and~\cite{kurzidim2009}) and other
features of the kinetic diagrams, one can see that even using our
custom routine the system setup represents a challenge to access the
state points of interest.

Note that in practice a probability of 50\% is unsatisfactory.
Ideally, none of the setup attempts should be rejected since as little
bias as possible should be exerted on the statistics. However, already
at relatively low $\Phim$ achieving $P = 100\%$ is strictly
impossible, the reason being that there exist peculiar matrix
configurations that prohibit all fluid particles to be inserted.
Nonetheless, by prolonging the CPU time $\theta$ it is possible to
increase $P$ since more theoretically possible system configurations
can actually be realized. Throughout this paper only state points have
been considered for which $P> 90\%$ was achieved.
\section{Finite-size effects}%
\label{apx:Finite-size.effects}%
\begin{figure}%
\includegraphics* [width = \Figurewidth]{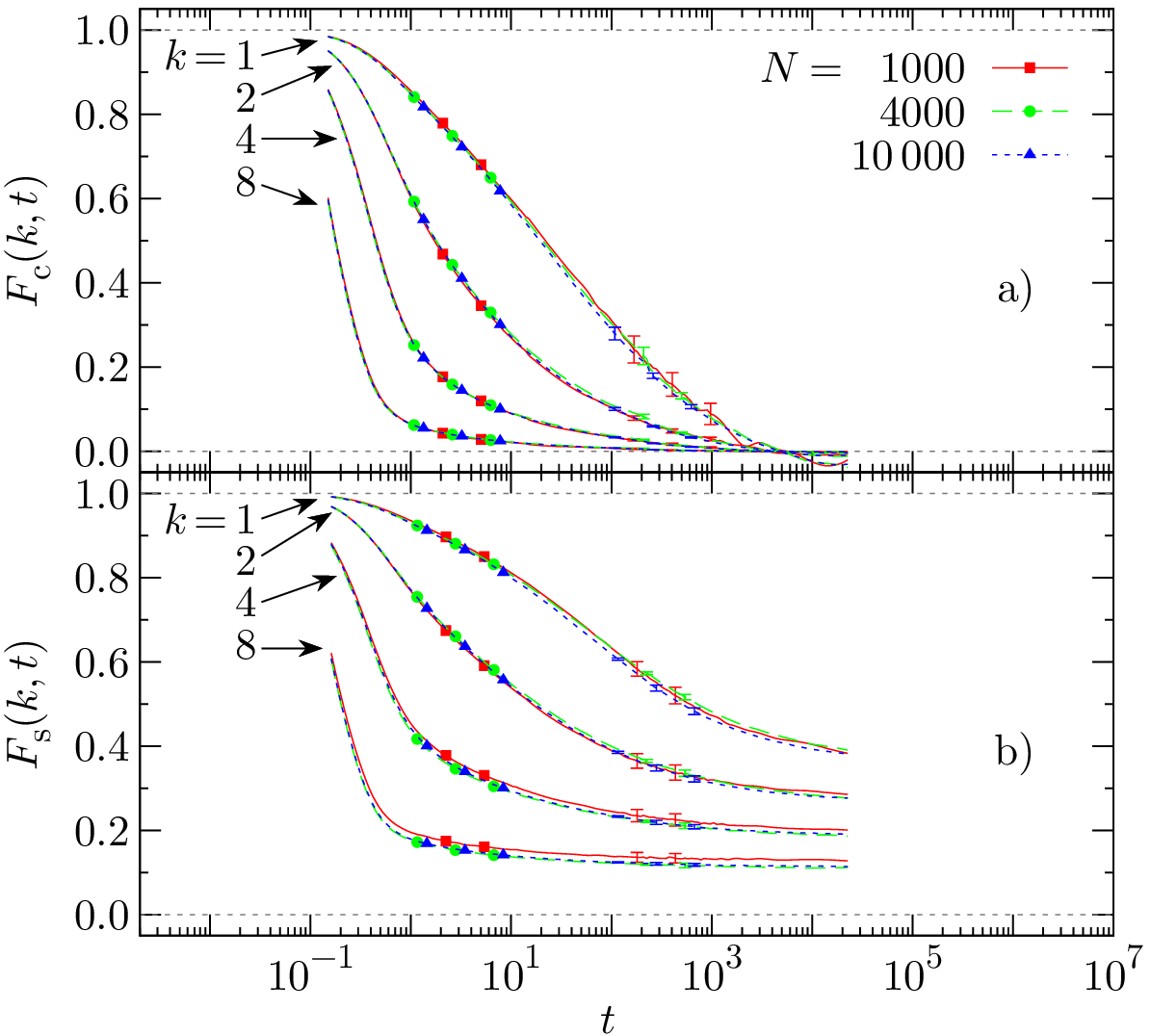}%
\caption{\label{fig:finite_size/fkt}%
(Color online) Assessment of the finite-size effects present at
$\Phim = 0.2500$ and $\Phif = 0.0125$. (a) $\Fc(k,t)$ and (b)
$\Fs(k,t)$, both displayed for $k \in \{ 1, 2, 4, 8 \}$ as well as
$N = 1000$ (red solid line, squares), $N = 4000$ (green dashed
line, circles) and $N = 10\,000$ (blue dotted line, triangles).}%
\end{figure}%
\begin{figure}%
\includegraphics* [width = \Figurewidth]{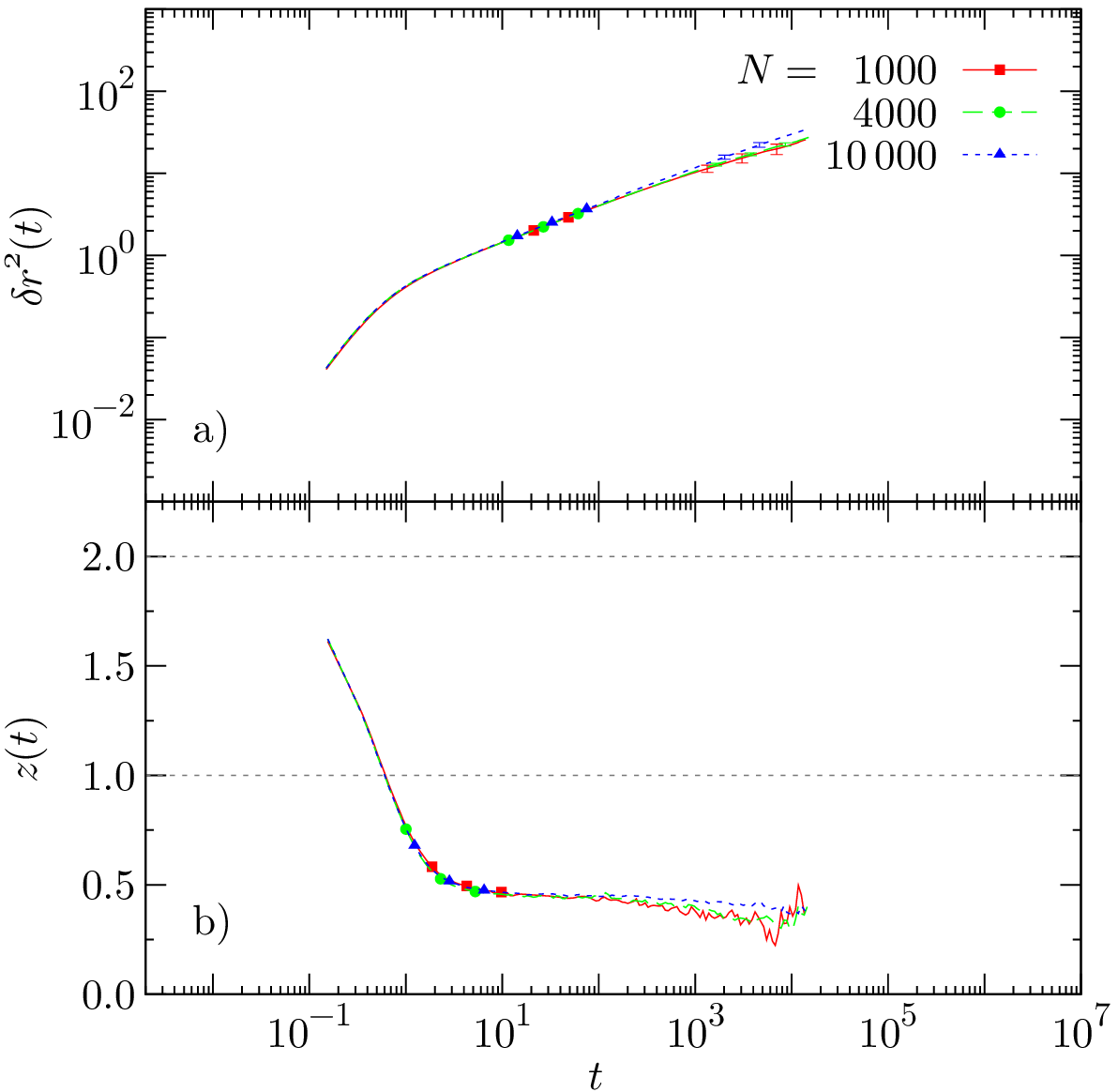}%
\caption{\label{fig:finite_size/msd}%
(Color online) Assessment of the finite-size effects present at
$\Phim = 0.2500$ and $\Phif = 0.0125$. (a) $\Drs(t)$ and (b)
$z(t)$, both displayed for $N = 1000$ (red solid line, squares),
$N = 4000$ (green dashed line, circles) and $N = 10\,000$ (blue
dotted line, triangles).}%
\end{figure}%

As mentioned in \prettyref{sec:Model.and.Methods}, throughout this
work we chose to assign the same constant total number of particles $N
= \Nm{+}\Nf = 1000$ to most systems simulated. The purpose of this
appendix is to determine for which state points this number is \emph{not}
sufficient to rule out the presence of finite-size effects, and
to describe our pertinent solution.

Strictly adhering to the approach of constant $N$, state points with a
large disparity of $\Phim$ and $\Phif$ attain low values of either
$\Nm$ or $\Nf$. Neither is desirable since it will render the
calculation of properties pertaining to either the matrix or the fluid
inaccurate. In this work, the most extreme case for low $\Nm$ is the
state point ($\Phim = 0.050$, $\Phif = 0.505$), with $N = 1000$
entailing $\Nm = 90$. The corresponding extremal case for $\Nf$ is the
state point ($\Phim = 0.3000$, $\Phif = 0.0025$), leaving only \emph{eight}
fluid particles to move in a matrix of $992$ particles. Hence,
in the present work the much more severe problem lies in low values of
$\Nf$.

To remedy this problem we chose to adjust $N$ such that neither $\Nm$
nor $\Nf$ assume a value below a lower limit of $N^* = 50$. The
previous brief analysis shows that in none of our systems $\Nm<N^*$,
which limits the discussion to state points with low $\Nf$. In the
following, we assess whether or not the choice of $N^*$ is reasonable.

In order to limit the effort, we consider a single state point
representative of the problem of low $\Nf$. Specifically, we inspect
various observables at $\Phim = 0.2500$ and $\Phif = 0.0125$, \Ie, at
a state point close to both $\Phif = 0$ and the percolation transition
in the voids (\Cf\ \prettyref{subsec:Path.III..constant.fluid.density} and
\ref{subsec:Trapping}). Using $N = 1000$ this system attains $\Nf = 48$ which
is
reasonably close to $N^*$. To investigate finite-size effects we
computed $\Fc(k,t)$, $\Fs(k,t)$, $\Drs(t)$, and $z(t)$ for $N = 1000$,
$4000$, and $10\,000$ while in each case averaging over ten matrix
configurations.

In \prettyref{fig:finite_size/fkt} the dependence of $\Fc(k,t;N)$
and $\Fs(k,t;N)$ upon $N$ is visualized. The intermediate scattering
functions are shown for selected wave vectors, where any finite-size
effects should be most prominent at small $k$. Indeed, for $k = 1.0$
and $t> 10^1$ there are noticeable discrepancies between $\Fc(k,t;
N{=}1000)$, $\Fc(k,t; N{=}4000)$, and $\Fc(k,t; N{=}10\,000)$;
however, this is hardly surprising considering that even for $N =
10\,000$ the corresponding length scale $2\pi/k = 2\pi$ is only
slightly smaller than half the edge length of the simulation cell. In
fact, already for $k = 2$, there is no distinguishable difference
between $\Fc(k,t; N{=}1000)$, $\Fc(k,t; N{=}4000)$, and $\Fc(k,t;
N{=}10\,000)$. On the other hand, $\Fs(k,t;N)$ displays small
consistent discrepancies between $N = 1000$, $4000$, and $10\,000$;
however, the worst deviation between $\Fs(k,t; N{=}1000)$ and
$\Fs(k,t; N{=}10\,000)$ (which is at $k = 8.0$ and $t = t_\mathrm{max}
= 2.25{\times}10^4$) is no more than ${\sim}12\%$. It is not
surprising that the $\Fs(k,t;N)$ curves in this case exhibit larger
deviations from one another than the $\Fc(k,t;N)$ curves considering
that the latter approach zero rather quickly whereas $\Fs(k,t;N)$ does
not completely relax. Also, we point out that any discrepancies in
$\Fs(k,t;N)$ for different $N$ values are well within the range of the
error bars.

Similarly, \prettyref{fig:finite_size/msd} illustrates the
dependence of $\Drs(t; N)$ and $z(t; N)$ upon $N$. The mean-squared
displacement displays only minor differences for $N = 1000$, $4000$,
and $10\,000$: The discrepancy between $\Drs(t; 1000)$ and $\Drs(t;
10\,000)$ at $t = t_\mathrm{max}$ amounts to ${\sim}24\%$; however,
since throughout this work we are interested in $\Drs(t;N)$ on \emph{logarithmic}
scales, the above deviation can be considered
sufficiently accurate. $z(t;N)$, on the other hand, is clearly more
susceptible to stochastic errors than $\Drs(t;N)$, leading those
errors to dominate finite-size effects in $z(t{>}3.6{\times}10^3;
N{=}1000)$. For $t = 3.6{\times}10^3$, we find a difference of mere
$17\%$ between $z(t; N{=}1000)$ and $z(t; N{=}10\,000)$, which permits
almost quantitative interpretation.

We conclude that $\Nf = N^*$ is sufficient to yield reliable results
for all observables considered. None of the latter exhibited a
deviation of more than $20\%$ between $N = 1000$ and the tenfold value
for $N$, which validates all conclusions presented in this work.

\end{document}